\documentclass[preprint]{elsarticle}
\usepackage{bm}
\usepackage{graphicx}
\usepackage{amsmath}

\renewcommand{\k}{{\kappa}}

\begin{document}

\begin{frontmatter}

\title{Exactly solvable spin-glass models with ferromagnetic couplings:
the spherical multi-$p$-spin  model in a self-induced field}
\author{Andrea Crisanti$^{1,2}$}

\author{Luca Leuzzi$^{1,3}$} 

\address{$1$ Dipartimento di Fisica, Universit\`a di Roma
``La Sapienza'', P.le Aldo Moro 5, 00185 Roma, Italy}

\address{$2$ CNR-ISC, Via dei Taurini 19, 00185 Rome, Italy}

\address{$3$ CNR-IPCF, UOS Roma, P.le Aldo Moro 5, 00185 Rome, Italy}

\begin{abstract}
We report some results on  the quenched disordered Spherical multi-$p$-Spin Model in
presence of ferromagnetic couplings. In
particular, we present the phase diagrams of some representative cases
that schematically describe, in the mean-field approximation, the behavior of most known transitions in 
glassy materials, including dynamic arrest in super-cooled liquids, amorphous-amorphous transitions and spin-glass transitions.
A simplified notation is introduced in order to compute systems properties in terms of 
an effective, self-induced, field encoding the whole ferromagnetic information.
\end{abstract}

\end{frontmatter}

\section{Introduction}

In the very extended framework of complex systems, spin glasses
have become the source of ideas and techniques now
representing a valuable theoretical background in diverse fields, with
applications far beyond the physics of amorphous materials (both
magnetic and structural).
These systems are characterized by a strong dependence from
the details, so strong that their behavior cannot be rebuilt starting
from the analysis of a single cell constituent. Their analysis cannot be
carried out without considering the collective behavior of the whole
system.  One of the common features is the occurrence of a large
number of stable and metastable states or, in other words, a large
choice in the possible realizations of the system. This goes along
with a rather slow evolution through many, detail-dependent
intermediate states, looking for a global equilibrium state (or
optimal solution).  Mean-field models have largely helped in
comprehending many of the mechanisms yielding such complicated
structure and also have produced new theories or combined among each
other old concepts pertaining to other fields such as, e.g., the
spontaneous breaking of the replica symmetry and the ultrametric
structure of states. Among mean-field models, spherical models - i.e., 
continuous dynamic variables with a global constraint \cite{Berlin52}- are
analytically solvable even in the most complicated cases. 

Multi-$p$-spin spherical models have been shown to yield low
temperature amorphous phases that, depending on the dominant
interaction terms, can both be described by discontinuous and
continuous Replica Symmetry Breaking (RSB) Ans{\"a}tze.  In
particular, (i) one step replica symmetry breaking (1RSB) phases were
studied, because of their relevance for the structural glass
transition \cite{Kirkpatrick87b,
  Kirkpatrick87c,Thirumalai88,Crisanti92}, (ii) two step RSB phases
\cite{Crisanti07b,Crisanti11} are found that are thermodynamically
stable and whose dynamics  models secondary relaxation in
glass-forming liquids (see, e.g., \cite{Romanini12} and for a thorough overview 
\cite{Ngai11} and references
therein) and study the singularities in the phase diagrams predicted
by the mode coupling theory \cite{Goetze89b,Goetze89c,Goetze09}, (iii)
the Full RSB phase represents spin-glasses in the proper sense and,
more generally, the frozen phase in random manifold problems
\cite{Mezard91,Giamarchi94, Giamarchi95,
  Cugliandolo96,LeDoussal98}. The
possibility of the existence of Full RSB in spherical models was first
pointed out by Nieuwenhuizen \cite{Nieuwenhuizen95} on the basis of
the similarity between the replica free energy multi-spin models and
the relevant part of the free energy of the Sherrington-Kirkpatrick
model.
 In Refs. \cite{Crisanti04b,
  Crisanti06}  thermodynamic stable Full RSB phases have been actually computed and analyzed.  
  Spherical models, thus, also
provide a much simpler realization of this Ansatz than in the spin-glass mean-field 
prototype model, i.e.,
the Sherrington-Kirkpatrick model \cite{Sherrington75}.

Further including ordered interaction terms representing attractive ferromagnetic 
couplings between spins, 
one can use
these models to study diverse problems, such as disordered systems along the Nishimori line
\cite{Nishimori11,Krzakala11}, or the states following
problem\cite{Barrat97,Capone06,Sun12}, else the random pinning with a system at
a very high temperature, or in presence of external random constraints, as, e.g.,  in porous media
\cite{Thalmann00,Krakoviack07,Krakoviack10}. Spherical models with 
competing disordered and ordered non-linear couplings  also describe mode-locking
laser models, where spherical spins  are used to represent both real and imaginary parts
of the complex amplitude of photonic modes  \cite{Gordon02,Gordon03, Weill05}. In
particular, they can be used to address the problem of random lasers
\cite{Cao99,Wiersma08}, whose statistical mechanics description involves
interactions between modes that are both non-linear and partially
quenched disordered \cite{Leuzzi09b,Conti11}.  In the latter case, we notice that the
global spherical constraint on continuous variables is not implemented
to approximate discrete spin variables or ease the computation of the
properties of continuous spins of fixed magnitude (like XY or
Heisenberg spins), but it represents the total amount of energy that
an external pumping laser beam forces into the random laser to activate
its modes.

We will show in this work that adding purely ferromagnetic terms to
the quenched disordered ones (a particular case of which is to have
quenched disorder with non-zero average) can be simply encoded into
adding an effective field to the purely disordered system. The paper
is organized as follow: in Sec. \ref{sec:model} we introduce the model 
and present a formal solution in the framework of Parisi Replica Symmetry Breaking Theory for the 
general case; in Sec. \ref{sec:s+p} 
we specialize the analysis to the $s+p$ case in an uniform external field; in 
Secs. \ref{sec:3+4} 
and \ref{sec:2+3} we study the behavior in presence of ferromagnetic couplings of two qualitatively 
different models both 
yielding RS and 1RSB phases: the $3+4$ and the $2+3$ models;
in Sec. \ref{sec:FRSB} we consider Replica Symmetry Breaking phases with continuous breakings 
and in Sec. \ref{sec:2+p}
we show an explicit case in which these phases appear, even in presence of competing 
ferromagnetic interactions. Eventually, in Sec. \ref{sec:unfolding} we show the termperature vs. degree of order  phase diagrams for the $2+p$ and the $3+4$ models, where the degree or order is yielded by a  combination of ferromagnetic
interaction magnitudes.
A word of caution. When the phase is described by a step-like order
parameter function, as the 1RSB phase, or it possesses a step-like part, the
transition between different phases can differ if one considers
the static or dynamic properties of the 
model \cite{Kirkpatrick87b,Kirkpatrick87c,Crisanti93,Crisanti07a}. 
When they are distinct one speaks of the static and dynamic
transitions. In the main text we shall  consider only the static transitions. 
The changes associated with the dynamic transition will be briefly discussed in \ref{App:B}.

\section{The Model}
\label{sec:model}
We consider the general model system described by the spin-Hamiltonian
\begin{equation}
\label{eq:model}
 {\cal H} = -\sum_{p\geq 2} \sum_{i_1<\cdots<i_p} \,
                    J^{(p)}_{i_1\cdots i_p}\,  \sigma_{i_1}\cdots \sigma_{i_p} 
                  -\sum_{k\geq 1} \frac{J^{(k)}_0}{N^{k-1}} \sum_{i_1<\cdots< i_k}
                  \sigma_{i_1}\cdots\sigma_{i_k}
\end{equation}
with both quenched, independently distributed, Gaussian $p$-spin interactions 
of zero mean and variance
\begin{equation}
\left[ \left(J^{(p)}_{i_1\cdots i_p}\right)^2\right] = \frac{p!\,J_p^2}{2 N^{p-1}},
\end{equation}
and uniform $k$-spin interactions $J^{(k)}_0$, with the $k=1$ term
representing the interaction with an external uniform field.  The
scaling of the interaction with the system size $N$ ensures the
correct thermodynamic limit $N\to\infty$.  The spins are real
continuous variables ranging from $-\infty$ to $+\infty$, subjected to
the global spherical constraint $\sum_{i}\sigma_i^2 = N$ that limits
the fluctuations and makes the partition function well defined.
The dynamics of the case  with a single $p>2$ term and $k=2$ was treated in Ref. \cite{Hertz99}.

The model can also be seen as  a spherical multiple-spin
interaction Spin Glass model with random couplings of {\sl non-zero}
average. The formulation (\ref{eq:model}) is, however, more general
since it gives more freedom in choosing the interactions in the
disordered and ordered part of the Hamiltonian. To stress this point
we have deliberately used different indexes, namely $p$ and $k$, for
the disordered and ordered interactions.

In the present study we shall consider the sub-class of models where
only two terms, one with $s$ and one with $p > s$ interactions, are
retained in the disordered part. These models have been called
spherical $s+p$ models \cite{Crisanti04b,Crisanti06,Krakoviack07b,Crisanti07c,Crisanti07b}. 
The hallmark of these models are  different
phase diagrams depending on the values of $s$ and $p$.  Representative
values of $s$ and $p$ will be discussed when needed.

\subsection{The partition sum and replicas}
\label{ss:partition}

The static properties of the model are obtained from the free energy
computed for fixed interactions and then averaged over the
disorder. This {\sl quenched} free energy can be computed using the
{\sl replica trick}: one first computes the {\sl annealed} free energy
density $\Phi(n)$ of $n$ non-interacting identical {\sl replicas} of
the system by rising the partition sum
  $Z= \mbox{\rm Tr}_{\sigma}\, e^{-\beta{\cal H}}$
to the $n$'th power and averaging it over the disorder:
\begin{equation}
 \Phi(n) = -\lim_{N\to\infty} \frac{1}{\beta N n} \ln \left[Z^n\right].
\end{equation}
The {\sl quenched} free energy density $\Phi$ is then obtained from the continuation of $\Phi(n)$ 
to non-integer values of $n$ down to $n=0$:
\begin{equation}
  \Phi = - \lim_{N\to\infty}\,\lim_{n\to 0} \frac{1}{\beta N n}\left( \left[Z^n\right] - 1\right)
          = \lim_{n\to 0} \Phi(n).
\end{equation}
In the last equality we assumed that the thermodynamic limit $N\to\infty$ and the replica limit
$n\to 0$ can be exchanged.
The calculation of $[Z^n]$ is rather standard, so we report the main steps, 
just in order to introduce our notation.
The interested reader can find more details in Refs. \cite{Mezard87,Crisanti92}.
By introducing the collective variables 
\begin{equation}
  q_{ab} = \frac{1}{N}\sum_{i}\sigma_i^a\,\sigma_i^b, \qquad
  m_a = \frac{1}{N}\sum_i \sigma_i^a
\end{equation}
where $a,b = 1,\ldots, n$ are {\sl replica indexes}, with $q_{aa} = 1$
from the spherical constraint, the leading contribution to $[Z^n]$ for
$N\to\infty$ can be written as
\begin{equation}
  \left[Z^n\right] \sim \int\, {\cal D}[q,\lambda, m, y]\, e^{- N G[q,\lambda, m, y]}, 
  \qquad N\to\infty
\end{equation}
where ${\cal D}[q,\lambda, m, y]\propto \prod_{a < b} dq_{ab}\, \prod_{a\leq b} \lambda_{ab}
                                                          \prod_a dm_a \prod_a dy_a$
                                                           denotes integrations over all (free) variables
and 
\begin{eqnarray}
 G[q,\lambda,m,y] &=& -\frac{1}{2}\sum_{ab} g(q_{ab}) - \sum_a \k(m_a) 
                                  + \frac{1}{2}\sum_{ab} \lambda_{ab}\, q_{ab} + \sum_a\, y_a\,m_a
                                  \nonumber
\\
                                  	                                  &&
                                  + \frac{1}{2}\ln\mbox{Det}\,(-\lambda)
                                  + \frac{1}{2}\sum_{ab}y_a (\lambda^{-1})_{ab}\, y_b.
\end{eqnarray}
We have introduced the short-hand notation:
\begin{equation}
 g(q) = \sum_{p\geq 2} \frac{\mu_p}{p}\,q^p, \qquad 
 \mu_p = \frac{p}{2} \beta^2 J_p^2
\end{equation}
\begin{equation}
  \k(m) = \sum_{k\geq 1} b_k\, m^k, \qquad 
   b_p = \frac{\beta}{k!} J^{(k)}_0
   \label{kappa}
\end{equation}
In the thermodynamic limit $N\to\infty$ the integrals can be evaluated
using the saddle-point approximation, leading to
\begin{equation}
\label{eq:phi}
 \beta \Phi = \lim_{n\to 0}\frac{1}{n}\, \mbox{Extr}\,G[q,m]
\end{equation}
with
\begin{equation}
  G[q,m] = -\frac{1}{2}\sum_{ab} g(q_{ab}) - \sum_a \k(m_a) 
                  - \frac{1}{2}\mbox{Tr}\ln\left(q_{ab} - m_a m_b\right)
\end{equation}
The functional $G[q,m]$ must be evaluated at its stationary point
that, as $n\to 0$, gives the maximum with respect to variations of
$q_{ab}$ and the minimum for variations in $m_a$.  The variables
$\lambda_{ab}$ and $y_a$ have been eliminated via the stationary point
equations. In the expression (\ref{eq:phi}) we have not included a
constant term $\beta \Phi_0$ that comes from neglected $O(N)$
terms. This fixes the zero temperature value of energy and entropy, but it is
not relevant for the study of the phase diagram.

Since we are interested into the limit $n\to 0$, the expression of $G[q,m]$ can be simplified further
by noticing that
\begin{equation}
 \mbox{Tr}\ln\left(q_{ab} - m_a m_b\right) = \mbox{Tr}\ln q
                               - \sum_{ab} m_a (q^{-1})_{ab}\, m_b + O(n^2)
\end{equation}
so we arrive at the final expression
\begin{equation}
\label{eq:gqm}
  G[q,m] = -\frac{1}{2}\sum_{ab} g(q_{ab}) - \sum_a \k(m_a) 
                  - \frac{1}{2}\mbox{Tr}\ln q + \frac{1}{2}\, \sum_{ab} m_a (q^{-1})_{ab}\, m_b
                  + O(n^2).
\end{equation}
By imposing stationarity of $G[q,m]$ with respect to 
variations with respect to $m_a$ and $q_{ab}$ ($a\not=b$) we obtain
the stationary point equations: 
\begin{equation}
\label{eq:spema}
 b(m_a) = \sum_{b} (q^{-1})_{ab}\, m_b
\end{equation}
\begin{equation}
  \Lambda(q_{ab}) + (q^{-1})_{ab} - \sum_{c}(q^{-1})_{ac} m_c\,
                                                               \sum_{c}(q^{-1})_{bc} m_c = 0,
                                                               \quad a\not=b
\end{equation}
\begin{equation}
\Lambda(q) \equiv \frac{d g(q)}{dq} \qquad; \qquad b(m) \equiv \frac{d \k(m)}{dm)} .
\end{equation}

To solve these equations we observe that eq. (\ref{eq:spema})  can be inverted to give
\begin{equation}
 m_a = \sum_{b} q_{ab}\, b(m_b) . 
\end{equation}
If we retain only the $k=1$ term in $\k(m)$, then $b(m) = b_1$. The
equation becomes $m_a = b_1 \sum_{b} q_{ab}$ and $m_a$ does not
depend on the replica index $a$.  This remains true in the general
case because there are no explicit replica symmetry breaking
fields. The stationary point equation for the magnetization $m_a
\equiv m$ then becomes
\begin{equation}
\label{eq:spem}
  m  = b(m) \sum_{b=1}^n q_{ab} \qquad \forall a=1, \ldots, n 
  \end{equation}
and the stationary point equation for $q_{ab}$ 
\begin{equation}
\label{eq:speq}
  \Lambda(q_{ab}) + (q^{-1})_{ab} + b(m)^2 = 0
                                                               \quad a\not=b.
\end{equation}
Note that if we consider the value of $b(m)$ as given, that is $b(m) =
b$, Eq. (\ref{eq:speq})  reduces to that of the
model in an external uniform constant field $h = T b$. This a rather
important technical point because we can split-up the resolution of the
stationary point equation into two steps. First we solve
eq. (\ref{eq:speq}) assuming $b(m) = b$ as fixed. Next we look for $m$
solution of eq. (\ref{eq:spem}) such that $b(m) = b$. In the following
we will  generically refer to $b$  as ``field".

\subsection{Parisi Parametrization: Replica Symmetry Breaking}
\label{ss:rsb}
To solve the self-consistent stationary point equation (\ref{eq:speq})
an assumption on the structure of the overlap matrix $q_{ab}$ must be
done. As the $n$ replicas of the real system are identical, one may
reasonably assume that the solution should be symmetric under the
exchange of any pair of replicas. In the high temperature (or field)
case this holds true, and the solution is of the form
\begin{equation}
  q_{ab} = \delta_{ab} + (1 - \delta_{ab})\,q_0.
\end{equation}
This form of $q_{ab}$ is known as the Replica Symmetric (RS) solution. 

As the temperature (and field) decrease the symmetry under replica
exchange is spontaneously broken, and the overlap matrix becomes a
non-trivial function of the replica indexes. In this regime the RS
assumption is not valid and a more complex structure arises.
Following the parameterization introduced by Parisi \cite{Parisi79a,Parisi79b},
the overlap matrix $q_{ab}$ for $R$ steps of replica permutation
symmetry breaking -- called RSB solution -- is divided along the
diagonal into successive blocks of decreasing size $p_u$, with $p_0 =
n$ and $p_{R+1} = 1$, and elements given by:
 \begin{equation}
 \label{eq:qpar}
  q_{ab} = q_{a\cap b} =  q_u , \qquad u = 0, \ldots , R + 1
 \end{equation}
with $1 = q_{R+1} \geq q_R \geq\cdots\geq q_1 > q_0$.  In this
notation $u = a\cap b$ denotes the overlap between the replicas $a$
and $b$, and means that $a$ and $b$ belong to the same box of size
$p_u$ but to two distinct boxes of size $p_{u+1} < p_u$.

The case $R = 0$ gives back the RS solution, while the limiting case
$R\to\infty $ produces the solution called Full Replica Symmetry
Breaking (FRSB or $\infty$-RSB) solution \cite{Parisi79b,
  Parisi80}. In this limit $q_u - q_{u-1} \to 0$ for $u = 1,\ldots,
R$, and the matrix $q_{ab}$ is described by a continuous,
non-decreasing function $q(x)$, where, in the Parisi parameterization,
$x$ varies between $0$ and $1$.
Solutions with a finite value of $R$ are called $R$-RSB solutions
\cite{Parisi80,Crisanti92,Krakoviack07b,Crisanti07c,Crisanti07b}.  These solutions can be
described by a step-like function $q(x)$.
Mixed-type solutions, with both discontinuous $R$-RSB-type and continuous FRSB-type parts 
for some $x$ interval, are also possible \cite{Crisanti04b, Crisanti06}. 

Inserting the form (\ref{eq:qpar}) into the free energy functional $G[q,m]$, eq. (\ref{eq:gqm}), with $m_a = m$,
one obtains 
\begin{eqnarray}
  \frac{2}{n} G[q,m] &=& - g(1) - \sum_{u=0}^R (p_u - p_{u+1})\, g(q_u) - \ln\left(1 - q_R\right)
  \nonumber\\
    &\phantom{=}&
                                     - \sum_{u=1}^R \frac{1}{p_u}  \ln\frac{\hat{q}_u}{\hat{q}_{u+1}} 
                                     -\frac{q_0 - m^2}{\hat{q}_1} - 2 \k(m)
\end{eqnarray}
where $\hat{q}_u$ is the Replica Fourier Transform (RFT) of $q_{ab}$
\cite{Carlucci96, DeDominicis97}:
\begin{equation}
   \hat{q}_u = \sum_{v=u}^{R+1}\, p_v\,(q_v - q_{v-1}).
\end{equation}

The free energy functional can be conveniently expressed by
introducing the auxiliary function
\begin{equation}
 x(q) = p_0 + \sum_{u=0}^R (p_{u+1} - p_u)\, \theta(q - q_u)
\end{equation}
which gives the fraction of pair of replicas with overlap $q_{ab} \leq q$. In terms of $x(q)$ the functional 
$G[q,m]$  takes the form
\begin{equation}
\label{eq:Gqm}
 \frac{2}{n} G[q,m] = - \int_{0}^{1} dq\, x(q) \Lambda(q) - \int_{0}^{q_R} \frac{dq}{\chi(q)} - \ln(1 - q_R)
                                   +\frac{m^2}{\chi(0)} - 2 \k(m)
\end{equation}
where
\begin{equation}
\label{eq:chiq}
  \chi(q) = \int_{q}^{1}\,dq'\, x(q').
\end{equation}
Note that $\chi(q_u) = \hat{q}_{u+1}$ and, moreover, $\chi(q) = \chi(q_0)=\chi(0)$ for $0\leq q\leq q_0$ since 
$x(q) = 0$ for  $q\in[0,q_0]$.

The stationary point equations are obtained from the first variation
of the free energy functional $G[q,m]$ with respect to $x(q)$ and $m$:
\begin{equation}
 \frac{2}{n} \delta G[q,m] = - \int_{0}^{1}\, dq\, F(q) ~\delta x(q) - 2\left[b(m) - \frac{m}{\chi(0)}\right]\, \delta m
\end{equation}
where
\begin{equation}
  F(q) = \Lambda(q) - \int_{0}^{q} \frac{dq'}{\chi(q')^2} + \frac{m^2}{\chi(0)^2}
\end{equation}
and
\begin{equation}
  \delta x(q) = \sum_{u=1}^{R} \left[\theta(q - q_{u-1}) - \theta(q - q_u) \right]\, \delta p_u 
                        - \sum_{u=0}^{R} (p_{u+1} - p_{u}) \delta(q - q_u)\, \delta q_u.
\end{equation}
Stationarity of $G[q,m]$ with respect to variations of $m$, $q_u$ and
$p_u$ gives:
\begin{equation}
\label{eq:spm}
 m = \chi(q_0)\, b(m)
\end{equation}
\begin{equation}
\label{eq:spq}
 F(q_u) = 0, \qquad u = 0,\ldots, R
\end{equation}
\begin{equation}
\label{eq:spp}
 \int_{q_{u-1}}^{q_u}\, dq\, F(q) = 0, \qquad u = 1,\ldots, R.
\end{equation}

The function $F(q)$ is continuous, thus eqs. (\ref{eq:spq}) and
(\ref{eq:spp}) require that between any two successive pairs
$(q_{u-1}, q_u)$ there must be at least two extrema of
$F(q)$. Denoting these by $q^*$, the extrema condition $F'(q^*) = 0$
implies that
\begin{equation}
  \int_{q^*}^1\, dq\, x(q) = \frac{1}{\sqrt{\Lambda'(q^*)}}
\end{equation}
where the prime denotes the derivative with respect to the argument $q$. 
The function $x(q)$ is a non-decreasing function of $q$, and the left
hand side of this equation is, thus, a concave function. The solutions
to this equation, thus, depend from the convexity properties of
$1/\sqrt{\Lambda'(q)}$: in the region where it is concave a continuum
of solution can be found, while where it is convex only discrete
solutions exist.  In the first case we deal with a continuous solution
of the FRSB-type, while in the second case with a R-RSB-type solution.
If $1/\sqrt{\Lambda'(q)}$ changes concavity for different intervals of
$q$, we have a mixed-type solution.

In the above argument the presence of {\em the ordered part of the
Hamiltonian does not play any role, once encoded into $b(m)$}. In this way one can decouple the computation studying  the behavior of a model in a (self-induced) ``external" field 
apart from the relationship between the field and the magnetizations induced by the ferromagnetic couplings.

 The value of the field only
enters in setting the value of $q_0$, the lowest possible value of
$q(x)$.  
As a consequence, the value $b$ of $b(m)$ fine-tunes the range
$[q_0, q_R]$ where solutions of the stationary point equations must be
searched.  Since $q_0$ is an increasing function of $b$ the presence
of an effective field $b$ can only reduce the ``complexity"
of the solution found in absence of it. In particular, by increasing
the value of $b$ we can eventually force $q_0 = q_R$, that is a transition to the
RS solution. For larger value of $b$ only the RS solution exists.

If terms besides the $k=1$ term (the uniform external field) are
present in the ordered part of the Hamiltonian, cf. Eq. (\ref{kappa}), what we just said is
only part of the game. In this case, indeed, $b(m)$ is a function of
$m$, and we must consider the possible solutions to Eq. (\ref{eq:spm})
such that  
\begin{equation}
\label{eq:unfolding}
b(m) = b
\end{equation}
 to unfold the complete solution.  The
unfolding depends on the form of $b(m)$. Therefore, starting from the same
phase diagram expressed as function of $b$, different phase diagrams
can be produced in the coupling constants, depending on the actual
$b(m)$.  An explicit instance  of complete phase diagrams in $T$, 
$J^{(s,p)}$ and  $J_0^{(s,p)}$ can be found in Ref. \cite{Sun12}
and the cases $2+p$ ($p\geq 4$) and $3+4$ will be reported in Section \ref{sec:unfolding}.
 In the
forthcoming part of the paper we shall, instead, address the fate of the different type of
solutions as the value of the effective field is varied.
To illustrate the results, we shall study $s+p$ models in
presence of an external uniform field $b$ described by the stationary
point equations (\ref{eq:spem}), (\ref{eq:speq}).
Stability of the stationary point requires that the quadratic form
\begin{equation}
  -\sum_{ab} \Lambda'(q_{ab})\left(\delta q_{ab}\right)^2 + 
            \mbox{Tr }\left({\bm q}^{-1}\delta {\bm q}\right)^2,
\end{equation}
must be positive (semi)definite,
where $\delta q_{ab} = \delta q_{ba}$ is the fluctuation of $q_{ab}$
from the stationary point value.  

\section{The Spherical $s+p$ model in an uniform external field}
\label{sec:s+p}

The Spherical $s+p$ model is the particular model obtained from the
general Hamiltonian (\ref{eq:model}) in which one retains only two
terms with random $s$-spin and $p$-spin interactions and a ($k=1$) uniform external field in Eq. (\ref{eq:model}).
Without loosing in generality, we assume $s < p$ from now on. For this model
we have:
\begin{eqnarray}
 g(q) &=& \frac{\mu_s}{s} q^s + \frac{\mu_p}{p} q^p, \quad
 \Lambda(q) = \mu_s q^{s-1} + \mu_p q^{p-1}, 
 \\
 \k(m) &=& b m \nonumber
\end{eqnarray}

The phase diagrams in the plane $(\mu_p, \mu_s)$, i.e. for $b=\beta h=0$, are well known \cite{Crisanti06}. Depending on
the value of $s$ and $p$ different type of solutions can be found, of both FRSB and R-RSB type and
mixed. 
Since we are interested on the effect of the external field $b$ on these different phases,
it can be useful to use the following parametrization 
\begin{equation}
\label{eq:paramr}
  \Lambda(q) = \mu_p\, (r q^{s-1} + q^{p-1})
\end{equation}
where
\begin{equation}
  r =\frac{ \mu_s }{ \mu_p} = \frac{s}{p} \frac{J_s^2}{J_p^2}, \qquad 0\leq r < \infty
  \label{eq:def_r}
\end{equation}
gives the relative strength of the $s$ and $p$ interaction terms, and we use $\mu_p$ and $b$ as 
free parameter for given $r$.
The temperature, when needed,  
is computed as $T/J_p = \sqrt{p/(2\mu_p)}$.

\subsection{The RS Solution}
\label{ss:RS}
All models, regardless of the value of $s$ and $p$, for large enough
temperature  (i.e., small enough $\mu_s$ and $\mu_p$) present a RS
phase. The equation for the RS phase are obtained inserting
\begin{equation}
 x(q) = \theta(q - q_0) 
\end{equation}
into equations (\ref{eq:spm}), (\ref{eq:spq}) and
(\ref{eq:spp}), or into the functional (\ref{eq:Gqm}), then making it
stationary with respect to $m$ and $q_0$. In either cases one ends up
with:
\begin{equation}
 m = (1 - q_0)^2 b
\end{equation}
and
\begin{equation}
\label{eq:rseq}
  \Lambda(q_0) = \frac{q_0}{(1-q_0)^2} - b^2
\end{equation}
The RS phase remains stable as long as the relevant eigenvalue
$\Lambda_1$ of the fluctuations remains positive:
\begin{equation}
\label{eq:rsstab}
 \Lambda_1 =   -\Lambda'(q_0) + \frac{1}{(1-q_0)^2} \geq 0.
\end{equation}
Using the stationary point equation (\ref{eq:rseq}) one obtains the equivalent condition
\begin{equation}
\label{eq:rsstab1}
 \Lambda(q_0) - q_0 \Lambda'(q_0) + b^2 \geq 0.
\end{equation}
The equal sign defines the critical line on which the RS phase
ends. With the help of the parameterization (\ref{eq:paramr}) the
parametric equation of the critical line reads:
\begin{equation}
\label{eq:rsctl}
\left\{
   \begin{array}{cl}
    \mu_p &= {\displaystyle \frac{1}{(1-q_0)^2} \frac{1}{r(s-1) q_0^{s-2} + (p-1) q_0^{p-2}} } \\
   	\phantom{0} & \phantom{0}\\
    \mu_s &= r\, \mu_p \\
   	\phantom{0} & \phantom{0}\\
     b^2 &= {\displaystyle \frac{1}{(1-q_0)^2} 
                                     \frac{r(s-2) q_0^{s-1} + (p-2) q_0^{p-1}} {r(s-1) q_0^{s-2} + (p-1) q_0^{p-2}} }
      \end{array}
      \right. \qquad 0\leq q_0 \leq 1
\end{equation}
Depending on the values of $\mu_p$ and $\mu_s$, the curve may show points where
\begin{equation}
  \left. \frac{d \mu_p}{d b}\right|_{\mu_s/\mu_p} = 0,
\end{equation}
that is $d T / d h = 0$ in the $(h,T)$ plane.  When present, one of
such critical points occurs where the transition between the RS phase and
the RSB one changes from continuous to discontinuous. A
discontinuity of finite height appears in $q(x)$ at the transition
and the critical line (\ref{eq:rsctl}) stops there.

By using the parametric form (\ref{eq:rsctl}), the points where $d\mu_p / d b = 0$ correspond 
to the value of $q_0$ solution of 
\begin{equation}
\label{eq:rs1rsbcl}
  2\Lambda'(q_0) + (q_0 - 1) \Lambda''(q_0)  = 0.
\end{equation}
The largest solution $ 0\leq q_c < 1$ of this equation, when it exists, gives the critical point where
the line (\ref{eq:rsctl}) ends, and $q_0$  gets  restricted to $q_c \leq q_0 \leq 1$.
For $r=0$ we recover $q_c = 1 - 2/p$ , while it is $q_c = 1 - 2/s$ in the
opposite limit $r\to\infty$. This is the critical value of the {\sl
  pure} $p$-spin, or $s$-spin, spherical model.  Beyond this point one
must resort to a 1RSB Ansatz in order to obtain the expression for the
transition line.

\subsection{The 1RSB Solution}
\label{ss:1RSB}
The 1RSB phase is described by a function $x(q)$ of the form
\begin{equation}
 x(q) = x\,\theta(q - q_0) + (1-x)\,\theta(q-q_1) 
\end{equation}
with $x\in [0,1]$ and $0\leq q_0 < q_1 \leq 1$. By plugging this
expression into the free energy functional $G[q,m]$ (\ref{eq:Gqm}),
and equating to zero its derivatives with respect to $q_0$, $q_1$, $x$
and $m$, or directly into the stationary point equations
(\ref{eq:spm}), (\ref{eq:spq}) and (\ref{eq:spp}), we obtain the 1RSB
(static) equations
\begin{equation}
\label{eq:1rsbq0}
  \Lambda(q_0) = \frac{q_0}{\chi(q_0)^2} - b^2
\end{equation}

\begin{equation}
\label{eq:1rsbq1}
  \Lambda(q_1) - \Lambda(q_0) = \frac{q_1 - q_0}{\chi(q_1) \chi(q_0)}
\end{equation}

\begin{equation}
\label{eq:1rsbx}
 g(q_1) -g(q_0) 
  - \left[
      \frac{q_0 - m^2}{\chi(q_0)^2} - \frac{1}{x} \frac{1}{\chi(q_0)}
     \right]
    + \frac{1}{x^2} \ln\left[\frac{\chi(q_1)}{\chi(q_0)}\right] = 0
\end{equation}
and
\begin{equation}
  m = \chi(q_0) b
  \label{eq:mvsb}
\end{equation}
where 
\begin{equation}
 \chi(q_1) = 1 - q_1, \quad 
 \chi(q_0) = 1 - q_1 + x (q_1 - q_0).
\end{equation}
For the purpose of the (numerical) solution of these equations, it is convenient to transform
eq. (\ref{eq:1rsbx}) into the equivalent expression
\begin{equation}
\label{eq:1rsbz}
 2\, \frac{g(q_1) - g(q_0) - \Lambda(q_0) (q_1 - q_0) }
            {(q_1 - q_0) \left[ \Lambda(q_1) - \Lambda(q_0) \right] } = z(y) 
\end{equation} 
 where $y = \chi(q_0) / \chi(q_1) \in [0,1]$  and  
 \begin{equation}
  z(y) = - 2 y \frac{1 - y - \ln y}{(1-y)^2}
 \end{equation}
 is the CS $z$-function \cite{Crisanti92}.  The advantage of equation
 (\ref{eq:1rsbz}) over (\ref{eq:1rsbx}) is that it does not depend on
 temperature. With the parameterization
 (\ref{eq:paramr}) it depends only on the ratio $r=\mu_s/\mu_p$. We
 can then easily solve the 1RSB equations for fixed $r$, and $x$.  The
 procedure is quite standard. One first introduces the ratio $t = q_0
 / q_1 \in [0,1]$ to be used as free parameter, and rewrite
 \begin{equation}
   q_1 = \frac{1 - y}{1 - y + x y ( 1 - t)}, \quad
   \chi(q_0) = \frac{x (1 - t) }{1 - y + x y (1-t)}.
 \end{equation}
 and $q_0 = t q_1$, $\chi(q_1) = y \chi(q_0)$. With this replacements
 eq. (\ref{eq:1rsbz}) becomes function of $y$ and $t$, besides $r$ and
 $x$.  Next one fixes the values of $r$ and $x$, and solve equation
 (\ref{eq:1rsbz}) for $y$ by varying $t \in [0,1]$. In this way one
 obtains $y \equiv y(t; x, r)$, that plugged into
 eqs. (\ref{eq:1rsbq1}) and (\ref{eq:1rsbq0}) gives the corresponding
 $\mu_p(t;x,r)$ and $b(t;x,r)$. This procedure builds the $x$-line for
 the 1RSB phase in the space $(\mu_p, \mu_s, b)$ on the plane $r =
 \mu_s / \mu_p$.
 
 The $x$-line with $x=1$ plays a special role. This line is the
 critical line separating the 1RSB and RS phases. The transition is
 discontinuous in $q(x)$ along the whole line since $t = q_0 / q_1 < 1$,
 and becomes continuous only at the end point $t\to 1$.
 
 The stability analysis of the 1RSB stationary point shows that the 1RSB solution is
 stable provided the eigenvalues
 \begin{equation}
   \Lambda_1^{(1)} = -\Lambda'(q_1) + \frac{1}{\chi(q_1)^2}
 \end{equation}
 \begin{equation}
  \label{eq:eig03}
   \Lambda_0^{(3)} = -\Lambda'(q_0) + \frac{1}{\chi(q_0)^2}
 \end{equation}
are both positive.
  
To illustrate the phases and the transition we shall now consider some explicit examples.

\section{The $3+4$ model}
\label{sec:3+4}
\begin{figure}[t!] 
   \centering
 \includegraphics[height=.99\textwidth, angle=270]{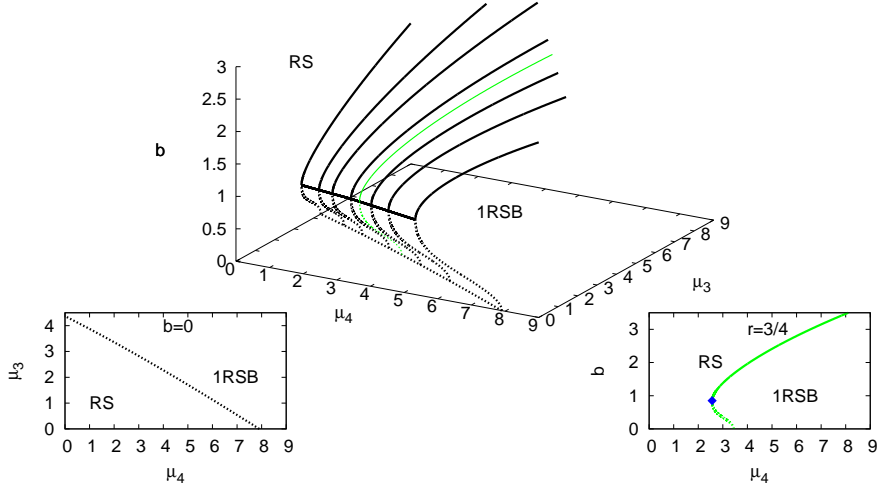} 
   \caption{Phase diagram of the $3+4$ model in the $(\mu_4, \mu_3,
     b)$ space (center).  The lines are the critical lines drawn for
     different values of $r$. Full black line: continuous transition;
     Dashed black line: discontinuous transition.  The discontinuous and
     continuous transition surfaces join on the end point line where
     $d\mu_3 / db|_r = 0$.  Left inset: projection on the $b=0$
     plane. Right inset: projection on the $r= 3/4$ plane (grey lines, green online); the dot
     is the end point $d\mu_4 / db|_r = 0$.}
     \label{fig:pd34_multi}
 \end{figure}

This model is the prototype of a system with only RS and 1RSB phases
separated by a discontinuous transition line. Indeed for $s=3$,
$p=4$ and $b=0$ eq. (\ref{eq:rsstab1}) reduces to
\begin{equation}
  -\mu_3 q_0^2 - 2\mu_4 q_0^3 \geq 0
\end{equation}
that can be satisfied only for $q_0 = 0$. For $q_0 = 0$ the eigenvalue
$\Lambda_1$, eq. (\ref{eq:rsstab}), reduces to $\Lambda_1 = 1 > 0$,
and RS solution with $q_0= 0$ is stable everywhere for $b=0$
\cite{Crisanti06}.  However, for large enough $\mu_3$ and $\mu_4$ a
1RSB phase with a favorable free energy appears. The transition
between the two phases occurs along the $x$-line with $x=1$.  The bottom-left inset in Figure
\ref{fig:pd34_multi} shows the phase diagram of the model for $b=0$.
 
 The phase diagram in the full $(\mu_4,\mu_3, b)$ space is reported in 
Figure \ref{fig:pd34_multi}. The lines, drawn for different values of $r$, define the critical 
surface separating the RS phase from the 1RSB phase. The transition between the two phases
is continuous on the part of the critical surface spanned by the full lines. 
This means that the difference $q_1 - q_0$ vanishes when approaching this portion of  critical 
surface from the 1RSB side. The breaking parameter $x$ takes
a value between $0$ and $1$ depending on the intersection point.  The transition turns into a 
discontinuous transition on the dashed part of the critical surface. Here the difference $q_1 - q_0$ 
remains finite as the critical surface is approached from the 1RSB side, but $x=1$. 
Indeed, the 
discontinuous part of the critical surface is the surface spanned by the $x$-lines with $x=1$ 
for different $r$.
The two parts of the critical surface join each other 
along the end point line, where $d\mu_4/d b|_r = 0$.
Along this line
\begin{equation}
  q_0 = q_1 = q_c = \frac{1}{12}\left[ 3(1-r) + \sqrt{9 r^2 + 6 r + 9}\right], \qquad 
    r\in [0,\infty).
\end{equation}
and
\begin{equation}
 \mu_4 = \frac{1}{(1-q_c)^2} \frac{1}{2 r q_c + 3 q_c^2}
\end{equation}
 \begin{equation}
   b^2 = \frac{1}{(1-q_c)^2} \frac{r q_c + 2 q_c^2}{2 r q_c + 3 q_c^2}
 \end{equation}
while $\mu_3 = r\, \mu_4$

In the bottom-right inset of Figure \ref{fig:pd34_multi} we report a slice of the phase diagram taken for fixed $r= 3/4$, 
though the plot is generic for all $r$. It is similar to the phase diagram
of the {\sl pure} $p$-spin spherical model in an external field. Indeed, by varying $r$ we smoothy interpolate
between the {\sl pure} $3$-spin and  $4$-spin spherical model in a field.

The scenario just described  remains valid for all values of $p > s > 3$, provided the difference
$p-s$ is not to large.\footnote{When $s\ll p$ a 2RSB phase arises in the frozen phase \cite{Crisanti07b,Crisanti11}.}

\section{The $2+3$ model}
\label{sec:2+3}
\begin{figure}[t!]
   \centering 
\includegraphics[height=.99\textwidth, angle=270]{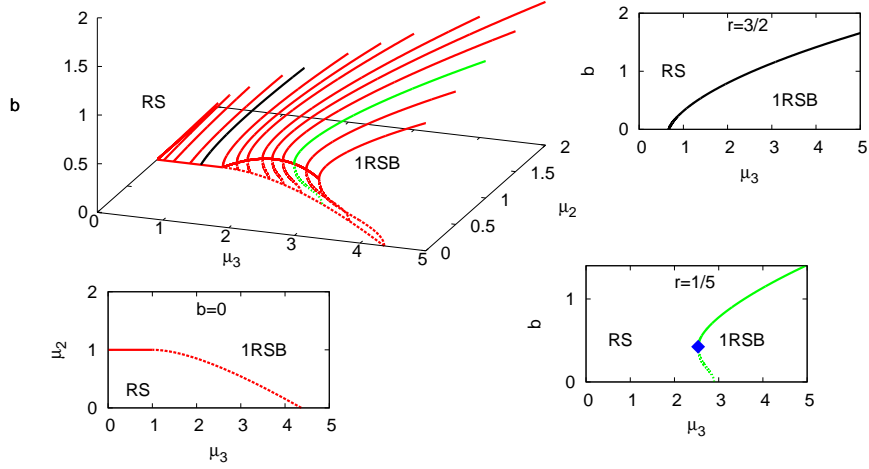} 
          \caption{Phase diagram of the $2+3$ model in the $(\mu_3, \mu_2, b)$ space.
                   The lines are the critical lines drawn for different values of $r$. 
                    Full line (red online): continuous transition; 
                    dotted line (red online): discontinuous transition; 
                    the end point line joining the discontinuous and continuous critical surfaces 
                    hits the $b=0$ plane at the point $(1,1,0)$. Bottom (left) inset: $b=0$ projection; the transition between the 
                    RS and the RSB phases is continuous along the full line, and discontinuous
                    along the dotted line. Two $x$-lines of the 1RSB are also shown. All $x$-lines
                    ends on the continuous transition line where $q_1=q_0=0$ and 
                     $\Lambda_1 = \Lambda_0^{(3)} =0$. Top (right) inset: projections on $r= 3/2$ (light grey full and dotted lines, green online) 
                     and $r= 1/5$ (black line) planes, from bottom to top.
                     For $r<1$ the discontinuous line merge into a continuous line (red dot). At $r=1$ the continuous transition
                  line hits the $\mu_3$ axis with an infinite slope. For $r>1$ no discontinuous transition
                   exists.
                    }
     \label{fig:pd23_mub}
\end{figure}

In the $3+4$ model the transition between the RS and 1RSB phases at small fields is always discontinuous.
Next in phase diagram complexity sits the $2+3$ model. For 
$b=0$ this model posses a phase diagram with only RS and 1RSB phases but, at difference 
with the $3+4$ model, the transition can be either continuous or discontinuous, 
see the bottom-left  inset of Figure \ref{fig:pd23_mub}.

When $s=2$, and $p$ generic, the stability condition (\ref{eq:rsstab1})  for $b=0$ becomes
\begin{equation}
 - \mu_p (p-2) q_0^{p-1} \geq 0
\end{equation}
that again is satisfied only for $q_0 =0$. However, at difference with the $3+4$ case, the relevant 
eigenvalue now reads $\Lambda_1 = 1 - \mu_2$, and vanishes for $\mu_2 =1$. 
Along this line the RS phase ($\mu_2 < 1$) becomes unstable against a 1RSB phase 
($\mu_2 > 1$), and a continuous transition between the two phases occurs. The continuous transition line ends at the point $(\mu_3, \mu_2) = (1,1)$ where the line hits the 1RSB $x$-line with 
$x=1$ \cite{Crisanti04b,Crisanti06}.

Figure \ref{fig:pd23_mub} shows the phase diagram of the $2+3$ model in the space 
$(\mu_3, \mu_2, b)$. The RS and 1RSB phases are separated by a critical surface. The transition 
can be either continuous, on surface spanned by full lines,
 or discontinuous,
on surface spanned by dotted lines. 
The continuous and discontinuous parts of the critical surface are joint along the
end point line where $d \mu_3 / db |_r =0$.  Here 
\begin{equation}
  q_0 = q_1 = q_c = \frac{1 - r}{3}, \qquad r \leq 1
\end{equation}
and 
\begin{equation}
 \mu_3 = \frac{1}{(1-q_c)^2} \frac{1}{r + 2 q_c}
\end{equation}
 \begin{equation}
   b^2 = \frac{1}{(1-q_c)^2} \frac{q_c^2}{r + 2 q_c}
 \end{equation}
while $\mu_2 = r\, \mu_3$.

For $r > 1$  the transition between the RS and 1RSB phases can only take place continuously,
with $q_0-q_1 \to 0$ on the critical surface. In Figure \ref{fig:pd23_mub}, 
 we show slices of the phase diagram
on the planes of constant $r$ above and below the critical value of $r=1$.

\section{Full RSB}
\label{sec:FRSB}
\begin{figure}[t!]
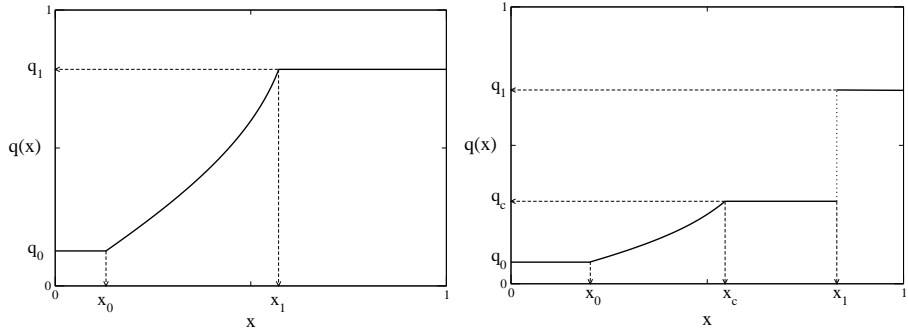
 
   \centering
   \includegraphics[width=.49\textwidth]{qx-frsb} 
   \includegraphics[width=.49\textwidth]{qx-1frsb} 
   \caption{Schematic form of the order parameter function $q(x)$ in the FRSB (left) and 1FRSB (right) phases. As  $q_c = q_1$ the two functions have equal form.
                   }
   \label{fig:qx-frsb}
\end{figure}

Moving to models more complicated than those mentioned above one can study the RSB in its continuous limit, as introduced by Parisi  to solve the Sherrington-Kirkpatrick model. That is, even in spherical models, one finds glassy phases whose correct thermodynamics can be computed only in this limit. 
The FRSB phase is described by a continuous order parameter function $q(x)$ of the form
\begin{equation}
  q(x) = \left\{
         \begin{array}{ll}
         q_0  = q(x_0)           &  0 \leq x \leq x_0   \\
         q(x)                            &  x_0 \leq x \leq x_1   \\
         q_1  = q(x_1)           &  x_1 \leq x < 1   
       \end{array}
\right.
\end{equation}
see Figure \ref{fig:qx-frsb},  solution of the 
stationary point equations
\begin{eqnarray}
\label{eq:qx0}
  \Lambda(q_0) &=& \frac{q_0}{\chi(q_0)^2} - b^2 \\
 \label{eq:qx}
  \Lambda(q) - \Lambda(q_0) &=&  \int_{q_0}^{q} \frac{dq'}{\chi(q')^2}, \qquad 
          q_0 \leq q \leq q_1
\end{eqnarray}
with $\chi(q)$  given by eq. (\ref{eq:chiq}).
To solve these equations we take the derivative of eq. (\ref{eq:qx}) with respect to $q$, leading to
\begin{equation}
\label{eq:lamp}
   \Lambda'(q) = \frac{1}{\chi(q)^2}, \qquad q_0 \leq q \leq q_1.
\end{equation}
By using this relation into eq. (\ref{eq:qx0}) we obtain the equation
\begin{equation}
\label{eq:q0frsb}
  b^2 = \frac{q_0}{\chi(q_0)^2} - \Lambda(q_0) 
         = q_0\, \Lambda'(q_0) - \Lambda(q_0)
\end{equation}
that solved for $q_0 = q_0(b)$ (or $b= b(q_0)$) fixes the lower bound of $q(x)$. 
To find the continuos part of $q(x)$ we observe that from the definition of $\chi(q)$, 
eq. (\ref{eq:chiq}), simply follows that $\chi'(q) = - x(q)$. As a consequence taking the derivative
of the relation (\ref{eq:lamp})  we have
\begin{equation}
\label{eq:qxfrsb}
 \Lambda''(q) = -\frac{2}{\chi(q)^3}\,\chi'(q) \quad \Rightarrow \quad
 x(q) = \frac{1}{2}\,\frac{\Lambda''(q)}{\left[\Lambda'(q)\right]^{3/2}},
 \qquad q_0 \leq q \leq q_1
\end{equation}
which gives the explicit analytic form of $x(q)$. Once inverted,  it
 leads to the continuous part of $q(x)$.
Inserting now the value of $q_0 = q_0(b)$ obtained from eq. (\ref{eq:q0frsb}) into $x(q)$ from 
eq. (\ref{eq:qxfrsb}) we have $x_0 = x(q_0) = x_0(b)$. 

To find the upper bound value $q_1$ of $q(x)$, and $x_1$, we observe that $\chi(q_1) = 1 - q_1$.
As a consequence, from eq. (\ref{eq:lamp}) evaluated for $q=q_1$ we obtain 
\begin{equation}
 \Lambda'(q_1) = \frac{1}{(1-q_1)^2}
 \label{eq:lambdap}
\end{equation}
that fixes the value of $q_1$. The value of $x_1$ follows   as $x_1 = x(q_1)$.

\subsection{The 1FRSB Solution}
\label{ss:1FRSB}
The 1FRSB solution differs from the FRSB for the presence of a discontinuous part in the
order parameter function $q(x)$:
\begin{equation}
  q(x) = \left\{
         \begin{array}{ll}
         q_0  = q(x_0)           &  0 \leq x \leq x_0   \\
         q(x)                            &  x_0 \leq x \leq x_c   \\
         q_c  = q(x_c)           &  x_c \leq x < x_1 \\
         q_1                           & x_1 \leq x < 1  
       \end{array}
\right.
\end{equation}
as schematically shown in the right side of Figure \ref{fig:qx-frsb}. 
The stationary point equations for the 1FRSB
are a ``mix" of those for the 1RSB and FRSB, and read:
\begin{eqnarray}
\label{eq:qx01}
  \Lambda(q_0) &=& \frac{q_0}{\chi(q_0)^2} - b^2 \\
 \label{eq:qx1}
  \Lambda(q) - \Lambda(q_0) &=&  \int_{q_0}^{q} \frac{dq'}{\chi(q')^2}, \qquad 
          q_0 \leq q \leq q_c
\\
  \Lambda(q_1) - \Lambda(q_c) &=& \frac{q_1 - q_c}{\chi(q_1) \chi(q_c)}
\end{eqnarray}
where, see eq. (\ref{eq:chiq}),
\begin{equation}
 \chi(q_1) = 1 - q_1, \quad 
 \chi(q_c) = 1 - q_1 + x_1 (q_1 - q_c)
\end{equation}
and 
\begin{equation}
  \chi(q) = 1 - q_1 + x_1 (q_1 - q_c)  + \int_{q}^{q_c}\, dq'\, x(q'), \qquad
   q_0 \leq q \leq q_c.
\end{equation}
The position of the breaking point $x_1$ follows from the equation
\begin{equation}
  2\, \frac{g(q_1) - g(q_c) - (q_1 - q_c) \Lambda(q_c)}
               {(q_1 - q_c) [\Lambda(q_1) - \Lambda(q_c)]} = z(y)
\end{equation} 
where $y = \chi(q_c) / \chi(q_1) \in [0,1]$. 

The 1FRSB solution reduces to the FRSB solution for $q_c = q_1$ or $x_1=1$.
In the former case the transition is continuous, and discontinuous in the latter.
When $q_0 = q_c$ the 1FRSB solution goes over a 1RSB solution.  

Similar to the FRSB solution, only $q_0$ (and $x_0 = x(q_0)$) depends on $b$. All 
other quantities remain unchanged by varying $b$. As a consequence, since $q_0$ grows
with $b$ (with $\mu_p$'s  held fixed) eventually 
 $q_0 = q_c$ and we observe a transition between the 1FRSB and the 1RSB phases.

\section{The $2+p$ model}
\label{sec:2+p}
\begin{figure}[t!] 
   \centering
   \includegraphics[height=.99\textwidth, angle=270]{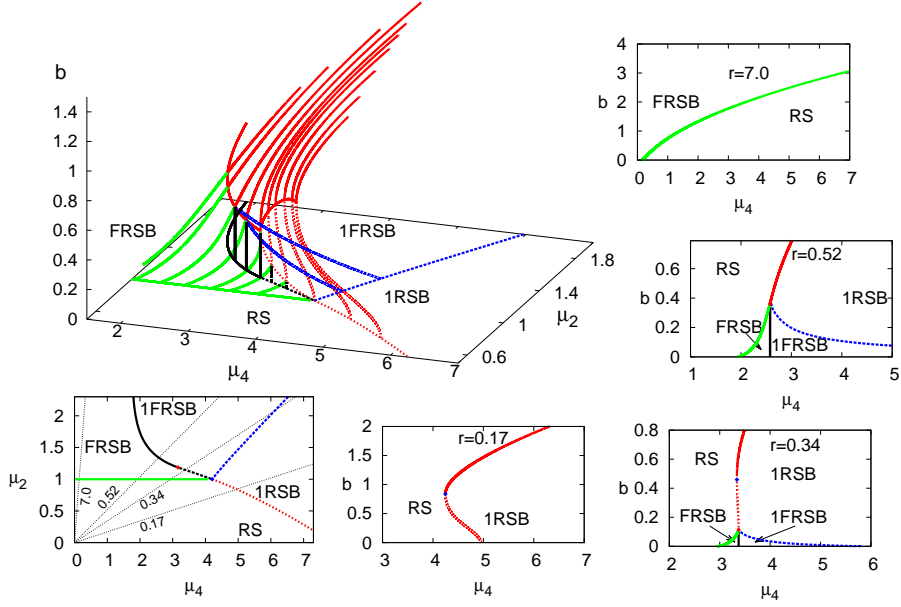} 
   \caption{Phase diagram of the $2+4$ model in the $\mu_2$, $\mu_4$, $b$ space. 
                   Dark grey full line (red online): RS-1RSB continuous transition.
                   Dark grey  dotted line (red online): RS-1RSB discontinuous transition.
                   Grey dashed line (blue online): 1RSB-1FRSB continuous transition.
                   Light grey full line (green online): RS-FRSB continuous transition.
                   Black full line: FRSB-1FRSB continuous transition.
                   Black dashed line: FRSB-1FRSB discontinuous transition.
                   Bottom Left:  $b=0$ projection. On the FRSB-1FRSB transition line (black) $q_c =  q_{1}$, 
                   while on the 1FRSB-1RSB 
                   transition line (dashed grey/blue) $q_c=0$. The transition line between the 
                   RS and the 1RSB phases 
                   (dark grey/red dotted) is
                   the $x$-line with $x=1$. The tiny dotted black lines on the $b=0$ plane denote different values of the
                    ($\mu_4$, $b$) plane at fixed $r=\mu_2/\mu_4$.
                    Each plot is generic for a given interval of $r$ values. 
                    Boundary values are shown in Tab. \ref{tab:r_values}.
                   From Top Right to Bottom Mid, clockwise: phase diagrams of the $2+4$ model in the plane $(\mu_4,b)$ with fixed $r=7.0, 0.52, 0.34$ and  $0.17$. 
                     }
   \label{fig:pd24_multi}
\end{figure}

The prototype model with both FRSB and 1FRSB phases is the $2+4$ spherical model. The model
belongs to the family of $2+p$ spherical models with $p>3$ whose phase diagram presents RS, 
1RSB, FRSB and 1FRSB phases. The phase diagram reproduced in the full $(\mu_4, \mu_2, b)$ space is shown
in Figure \ref{fig:pd24_multi}, where the four different phases and relative transition 
lines are indicated.
 Its 
$b=0$ projection is shown in the  bottom-left inset.
\begin{table}[b!]
\begin{center}
\begin{tabular}{|ll|}
\hline
Critical $r^{t}_x$ values &  \qquad Phases (transition kinds)  \\ \hline
&\\
 & \hspace*{-1cm} RS, FRSB (continuous)\\ 
$r_0^{(1)}=6$& \hspace*{-1cm} $-----$\\
 & \hspace*{-1cm} RS, FRSB, 1FRSB (continuous)\\ 
$r_0^{(0)}=0.6382$ &\hspace*{-1cm}  $-----$ \\
 & \hspace*{-1cm} RS, 1RSB, FRSB, 1FRSB (continuous)\\ 
$r_1^{(1)}=0.375$&\hspace*{-1cm}  $-----$\\ 
& \hspace*{-1cm} RS, 1RSB, FRSB, 1FRSB (cont. and disc.)\\  
$r_1^{(0)}=0.2378 $&\hspace*{-1cm} $-----$ \\ 
& \hspace*{-1cm} RS, 1RSB (cont. and disc.)\\ \hline
    \end{tabular}
\end{center}
\caption{Boundary values of $r=\mu_2/\mu_4$ between different kinds of $(\mu_4,b)$ phase diagrams. The top index in $r^{(t)}_x$ is the value of $t=q_0/q_1=0,1$, the sub-index is the value of $x=0,1$.
See Fig. \ref{fig:pd24_multi} for a graphical representation.}
\label{tab:r_values}
\end{table}

We now discuss the conduct  of each phase when the field $b$ is switched on.
We first analyze the FRSB and 1FRSB phases, 
then the RS phase, and, eventually, the 1RSB phase.

\subsection{$2+p$: FRSB phase}
The FRSB solution is known to reproduce the low temperature phase of the $2+p$ model at $b=0$ \cite{Crisanti04}. In this special case,  equations (\ref{eq:qx0})-(\ref{eq:qxfrsb}) become:
\begin{equation}
	q_0 = \left[\frac{b^2}{\mu_p (p-2)}\right]^{1/(p-1)}
	\label{eq:q0_b}
\end{equation}
\begin{equation}
 x(q) = \frac{1}{2}\, \frac{ \mu_p (p-1)(p-2) q^{p-3} }
                                         { \left[  \mu_2 + \mu_p (p-1) q^{p-2} \right]^{3/2} }
\end{equation}
\begin{equation}
 \mu_2 + \mu_p (p-1) q_1^{p-2} = \frac{1}{(1-q_1)^2}
\end{equation}

At $b>0$ the FRSB may exist only if the ratio $r=\mu_2/\mu_p$ is larger than the critical 
value $r_1^{(0)}$, cf. Tab. \ref{tab:r_values}.
When the field $b$ is switched on only $q_0$ is modified: 
it becomes non-zero and grows with $b^{2/(p-1)}$, c.f.r equation (\ref{eq:q0_b}). 
Increasing $b$ 
for fixed $\mu_2$ and $\mu_p$, we eventually have 
$q_0= q_1$  and a continuous transition  from the FRSB  to the RS phase takes place.
In Figure \ref{fig:pd24_multi} the RS-FRSB critical surface for the $p=4$ case 
is the one spanned by the light grey (green)  lines.
For any $(\mu_4,b)$ slice with $r > r_0^{(1)}$ the FRSB phase is bounded exclusively by the RS phase,
see top inset of Fig. \ref{fig:pd24_multi}.
At $r = r_0^{(1)}$ a boundary with the 1FRSB phase first appears, the relative critical surface 
is the one spanned by the black lines in Fig. \ref{fig:pd24_multi}. The transition can be either
continuous, if  $r_1^{(1)} < r < r_0^{(1)}$ (full continuous lines, see also the inset for $r=0.52$ in 
Fig. \ref{fig:pd24_multi}),
or discontinuous, if  $r_1^{(0)} < r < r_1^{(1)}$ (dashed lines, cf. inset of Fig. \ref{fig:pd24_multi} for $r=0.34$).

\subsection{$2+p$: 1FRSB phase}
The 1FRSB phase may exist only if $r_1^{(0)} < r < r_0^{(1)}$, cf. Tab. \ref{tab:r_values} The fate of the 1FRSB phase 
in a field is similar to that of the FRSB phase because the field $b$ only affects the
 value of $q_0\sim b^{2/(p-1)}$.
 As a consequence, for $b$ large enough, $q_0 = q_c$ and the 1FRSB phase goes over the 
 1RSB phase. The transition is clearly continuous.  The 1FRSB-1RSB surface is
 the one spanned by the grey (blue)  dashed lines in Fig. \ref{fig:pd24_multi} and insets.
The  other boundary of the 1FRSB phase is  with the FRSB phase. The transition  between these two 
phases can  be either discontinuous or continuous, depending on $r$ being smaller or larger than 
$r_1^{(1)}$, as we mentioned above, discussing  the FRSB phase. 
We note that for $b=0$ the 1FRSB phase is bounded by the 1RSB phase only when
$r_1^{(0)} < r < r_0^{(0)}$. See, in Fig. \ref{fig:pd24_multi} the insets for $r=0.52$ and $r=0.34$.

\subsection{$2+p$: RS phase}
The field $b$ has an ordering effect on the system, i.e.,  for large value of $b$ the RS phase 
is the stable phase, c.f.r. eq. (\ref{eq:rsstab1}).  
When $b$ is decreased the disordered terms in the Hamiltonian 
become more and more relevant and, depending on the value of $\mu_2$ and $\mu_p$,
the RS phase may become unstable towards a more
complex phase.  This happens when the relevant eigenvalue $\Lambda_1$ becomes negative, 
see eq. (\ref{eq:rsstab}).
The vanishing of $\Lambda_1$ defines the critical RS surface,  whose parametric 
equation is given by eq.
(\ref{eq:rsctl}) for $s=2$. In Figure \ref{fig:pd24_multi} for $p=4$ this surface is the one spanned by
the light grey (green online) and dark grey (red) full lines. If the value of $b$ is further decreased one enters into either
the FRSB phase, crossing the light grey (green) lines in figure,  or the 1RSB phase, 
crossing the full dark grey (red) lines in figure. 
The transition is in either case continuous.
Which phase may be reached as the field $b$ varies depends on the ratio $r$ 
between the coupling coefficients $\mu_2$ and $\mu_p$. For
$r  > r_0^{(1)}$ only the FRSB phase can be encountered,  see top inset of  Fig. \ref{fig:pd24_multi} for $p=4$,
while for $r < r_1^{(0)}$ only the 1RSB phase is feasible: see bottom inset of  Fig. \ref{fig:pd24_multi}. 
In between ($r_0^{(1)}> r>r_1^{(0)}$) both 1RSB and FRSB  are possible, see insets for $r=0.34$ and $r=0.52$ 
in Fig. \ref{fig:pd24_multi}.

The continuous RS critical surface may bend, and become multivalued, if 
 $d\mu_p/db=0$ for fixed $r$.  This occurs when $q_0$ equals the
largest solution $0 < q_c (r)< 1$ of
\begin{equation}
\label{eq:fq}
  f(q) = p(p-1) q^{p-2} - (p-1)(p-2) q^{p-3} + 2 r =0,
\end{equation}
c.f. eq. (\ref{eq:rs1rsbcl}) for $s=2$.
For  large enough $r$ this equation has no physical solutions. They do appear when $r$ is
sufficiently small. For $r=0$ we have $q_c(0) = (p-2)/p$.
To find the critical value of $r$ we observe that  
 $f(q)\sim - (p-1)(p-2) q^{p-3}$ for $q\simeq 0^+$ while
$f(q) \sim p(p-1) q^{p-2}$ for $q\gg 1$. Hence, there must be at least one minimum $f'(q^*)= 0$ with $q^* > 0$.
A simple calculation yields
$q^* = (p-3)/p$.
This, with $f(0) = 2 r$,  implies that, if $r$ is not too large,  equation  (\ref{eq:fq}) has at least two 
solutions with $q> 0$, and the surface may bend.
Since increasing $r$ shifts $f(q)$ upwards, a crossover value of $r$ is obtained by imposing 
that $f(q^*) = 0$. This leads to 
\begin{equation}
r_1^{(1)} = \frac{(p-1) (p-3)^{p-3} }{2 p^{p-3}}
\label{eq:r_1_1}
\end{equation}
The value $q_c(r)$ for $r <  r_1^{(1)}$ defines the end point line where the surface bends.
For the special case $2+4$ reported in Fig. \ref{fig:pd24_multi},  eq. (\ref{eq:fq}) can be explicitly solved and one finds
\begin{equation}
  q_c(r) = \frac{1}{12}\left[3 + \sqrt{9 - 24 r}\right],
  \qquad r < r_1^{(1)} = 3/8.
\end{equation}
When the continuous RS critical surface bends, the transition between the RS and the 1RSB 
phases becomes discontinuous on the end point line $q_c(r)$. 
This line is the dark grey (red) line joining the contact points between the full and dotted dark grey (red) lines
in Figure \ref{fig:pd24_multi}. 
Below this line the transition between the RS and 1RSB is discontinuous and occurs on the 
critical surfaces spanned by the $x$-line of the 1RSB phase with $x=1$. This  is represented by
 dark grey dotted lines in figure \ref{fig:pd24_multi} (see  insets  for $r=0.17, 0.34$).

We note that since the surface is bended there is  an inverse transition in $b$. That is, 
by decreasing $b$ we can enter the 1RSB phase from the RS phase via a continuous
transition and, decreasing further $b$, leave the 1RSB phase for the RS phase again via a discontinuous
transition.

\subsection{$2+p$: 1RSB phase}
The 1RSB phase is found for large enough $\mu_2$ and $\mu_p$, see e.g. the bottom-left
inset of Fig. \ref{fig:pd24_multi} for the $2+4$ model with $b=0$.  When $b$ increases $q_0$
grows and eventually becomes equal to $q_1$. Here the 1RSB phase ends.
Since as $q_0\to q_1$ the second 1RSB equation (\ref{eq:1rsbq1}) reduces to the critical 
condition $\Lambda_1=0$ of the RS phase, c.f.r eq. (\ref{eq:rsstab}), one enters 
into the RS phase through the continuous RS-1RSB transition, 
the surface spanned by the dark grey (red) full lines in Fig. \ref{fig:pd24_multi}.
This  critical surface  bounds the 1RSB from "above".

As discussed for the RS phase,  if the ratio $r$ between $\mu_2$ and $\mu_p$ is smaller than 
the critical value $r^{(1)}_1$  the critical surface $\Lambda_1=0$ bends. Where this happens
the transition between the 1RSB and RS phases takes place with a finite value of $q_1-q_0$
and occurs when we cross the surface spanned by the $x=1$ lines of the 1RSB solution.
In Fig. \ref{fig:pd24_multi} this is represented by the surface spanned by the dark grey (red)
dotted lines.

If the ratio $r$ exceeds the critical value $r^{(0)}_1$ the eigenvalue $\Lambda_0^{(3)}$
may become negative for low $b$, and the 1RSB phase is unstable with respect to 
a 1FRSB phase. The transition between the 1RSB and the 1FRSB phase occurs on the critical 
surface defined by $\Lambda_0^{(3)}=0$.

This critical surface intersects the $b=0$ plane along the critical line of equation, see \ref{App:A},
\begin{equation}
\label{eq:mup1rsb2pt0}
  \mu_p =  \frac{(1 - y_0 + x y_0)^p}{x^2 y (1-y_0)^{p-3}}
\end{equation}
\begin{equation}
\label{eq:mu21rsb2pt0}
  \mu_2 =  \frac{(1 - y_0 + x y_0 )^2}{x^2},
\end{equation}
where $y_0$ solution of 
\begin{equation}
\label{eq:2pzyt0}
 z(y_0) = \frac{2 + (p-2) y_0}{p},
\end{equation}
This is  shown as dashed grey (blue) lines  in Figure \ref{fig:pd24_multi}.
Along this line the ratio $r = \mu_2/\mu_p$ is given by
\begin{equation}
  r = \frac{ y_0 (1-y_0)^{p-3} }{ (1 - y_0 + x y_0)^{p-2} }
\end{equation}
and varies between
\begin{equation}
r_1^{(0)} = y_0 (1 - y_0)^{p-3}
\label{eq:r_1_0}
\end{equation}
 for $x=1$ 
 and
\begin{equation}
r_0^{(0)} = \frac{y_0 }{ 1 - y_0}
\label{eq:r_0_0}
\end{equation}
 for $x=0$.
For $r$ in this range  both FRSB and 1FRSB phases, beside the 1RSB phase, may exist,
see inset of Fig.  \ref{fig:pd24_multi} for $r=0.34$.

When $x=1$ the critical line intersects both the RS-1RSB and RS-FRSB critical lines at the
multi-critical point.
For  $r < r_1^{(0)}$  only the RS and 1RSB phases exist,
see inset of Figs \ref{fig:pd24_multi} for $r=0.17$.

In the opposite limit $x=0$ both $\mu_p$ and $\mu_2$ diverge, while its ratio $r$
remains finite. As a consequence for $r > r_0^{(0)}$
only a 1RSB phase with $q_0\not= 0$ may exist, provided $r < r_0^{(1)}$ (see below),
cf. $r=0.52$ inset of Fig. \ref{fig:pd24_multi}.
The numerical values for $p=4$ are:
$y_0 = 0.389571$,
$r_1^{(0)} = 0.2378$, $r_0^{(0)}= 0.63819$.

The breaking point $x$ cannot exceed $1$, implying that the 1RSB critical surface is bounded by 
the $x$-line with $x=1$ 
(dark grey/red dotted line in Figure \ref{fig:pd24_multi}).
This line intersects the $b=0$ plane at the end point of
the 1RSB-1FRSB critical line on the $b=0$ plane 
and gives the continuation to $b \not= 0$ of the critical 
RS-1RSB line found on the $b=0$ plane.

The other boundary of the 1RSB-1FRSB critical surface occurs when  
 $q_1=q_0$,  
and the discontinuity associated with the 1RSB (and 1FRSB) solution disappears.
The equation of the critical end line of the 1RSB-1FRSB critical surface reads, see \ref{App:A},
\begin{equation}
   \mu_p = \frac{2}{27}\,  
                   \frac{ (p - 3 + 3 x)^p}
                           {  (p-1)(p-2) (p-3)^{p-3} x^2}
\label{eq:mup_1rsb_rs}                           
\end{equation}
\begin{equation}
 \mu_2 =   \frac{p}{27} \, \frac{(p - 3 + 3 x)^2}{(p - 2) x^2}
\label{eq:mu2_1rsb_rs}                           
\end{equation}
\begin{equation}
  b^2 =  \frac{2}{27} \frac{(p - 3 )^2 (p - 3 + 3 x)}{ (p - 1) x^2}.
\label{eq:b2_1rsb_rs}                           
\end{equation}
By varying $x$ between $0$ and $1$ we obtain the critical line where 
the RS-1RSB-1FRSB-FRSB phases meet  altogether. This is represented as a 
dark grey/red full  line for $r > r_1^{(1)}$ in Figure \ref{fig:pd24_multi},  
see also the $r=0.52$ inset.
Along this line the ratio $r$ is given by
\begin{equation}
r = \frac{p(p-1)}{2} \, \frac{(p-3)^{p-3} } 
                                            {(p-3 + 3 x)^{p-2}}
\end{equation}
 varying between $r_1^{(1)}$, cf. Eq. (\ref{eq:r_1_1})
   for $x=1$,
and 
\begin{equation}
r_0^{(1)} = \frac{p ( p-1) (p-3)^{p-4} }{2}
\label{eq:r_0_1}
\end{equation}
 for $x=0$.

When $x=1$ the line meets the critical $x$-line with $x=1$. Hence,
a discontinuous RS-1RSB transition can be found only if $r < r_1^{(1)}$,
while the transition is always continuous for $r > r_1^{(1)}$,
as shown in the $(\mu_4,b)$ diagrams at fixed $r$  in 
Fig. \ref{fig:pd24_multi}.

As it occurs along the 1RSB-1FRSB surface, for $x=0$ both  $\mu_p$ and $\mu_2$, as well as $b$,  diverge. The ratio $r$, nevertheless,
remains finite.
As a consequence, the 1RSB phase cannot be found if $r > r_0^{(1)}$, and
only the FRSB phase survive, see Fig. \ref{fig:pd24_multi}.
For $p=4$,  $r_1^{(1)} = 3/8$ and $r_0^{(1)} = 6$.

\section{Phase diagrams of the $s+p$ model in the ferromagnetic interaction strength}
\label{sec:unfolding}
The spherical $s+p$ model with ferromagnetic interactions is the particular model described by
\begin{equation}
 \Lambda(q) = \mu_s q^{s-1} + \mu_p q^{p-1}, \qquad
 k(m) = b_s m^s + b_p m^p
\end{equation}
where, as usual, it is assumed $p>s$.

Next to equations (\ref{eq:paramr}-\ref{eq:def_r}) it is convenient to introduce the parameterization 
\begin{equation}
\label{eq:paramr2}
   \k(m) = b_p\, (\gamma m^s + m^p)
\end{equation}
where
\begin{equation}
  \gamma = \frac{b_s }{  b_p} = \frac{p!}{s!} \frac{J_0^{(s)}}{J_0^{(p)}}, \qquad 0\leq \gamma < \infty
\end{equation}
gives the relative strength of the $s$ and $p$ interaction terms, and use $\mu_p$ and $b_p$ as 
free parameter for given $r$ and $\gamma$. We do not consider the case of competing
ferro-antiferromagnetic interaction ($\gamma < 0$), but the extension is straightforward.

With this parameterization the ``unfolding equations" $b(m)=b$ and $m=\chi(q_0) b$, 
cf. equation (\ref{eq:unfolding}) and Sec. \ref{ss:rsb},
yield
\begin{eqnarray}
\label{eq:unfsp}
 b(m) &=& \frac{m}{\chi(q_0)} \quad \Rightarrow\quad
 \nonumber
 \\
   b_p &= &\frac{1}{\chi(q_0)} \frac{1}{\k'(m)/m}
           =\frac{1}{\chi(q_0)} \frac{1}{\gamma s m^{s-2} + p m^{p-2}}
\end{eqnarray}
which, with eqs. (\ref{eq:1rsbq0})-(\ref{eq:mvsb}), gives $b_p$ as function of SG parameters ${\mathbf q}=\{q_0,q_1,\ldots\}$ 
and ${\underline \mu}=\{\mu_p,\mu_s\}$
in the case of an uniform external field.

The  {\sl natural} parameters $\mu_p$ and $b_p$ can be transformed into the {\sl physical} 
parameters that give the temperature and the strength of the ferromagnetic  part by setting
\begin{equation}
  J_p^2 = \alpha\, J^2, \qquad 
  J_s^2 = (1-\alpha)\, J^2, \qquad 
  \alpha = \frac{s}{s + r p}
\end{equation}
and
\begin{equation}
  J_0^{(p)} = \alpha_0\, J_0, \qquad 
  J_0^{(s)} = (1-\alpha_0)\, J_0, \qquad 
  \alpha_0 = \frac{p!}{p! + r s!}
\end{equation}
where $J$ and $J_0$ measure the overall strength of the disorder and ferromagnetic parts.
One then has
\begin{equation}
  \mu_p = \frac{p}{2} \alpha \beta^2 J^2 \quad \Rightarrow \quad
  T/J = \sqrt{\frac{p\alpha}{2\mu_p}} = \sqrt{\frac{sp}{2(s + rp)\mu_p}} 
\end{equation}
\begin{equation}
  b_p = \frac{\alpha_0}{p!} \beta J_0 \quad \Rightarrow \quad
  J_0 = \frac{p!}{\alpha_0}T\, b_p = (p! + \gamma s!)T b_p.
\end{equation}

For fixed temperature $T$, or $\mu_p$, the Ferromagnetic (FM) solution $m\not=0$ first appears
at the critical value 
\begin{equation}
 J_0^* = \min_{q_0} J_0({\mathbf q},{\underline \mu}).
\end{equation}
For $J_0  < J_0^*$ only the Paramagnetic (PM) solution $m=0$ is possible.

The PM/FM transition can be either continuous or discontinuous in $m$. 
If the minimum of $J_0$ occurs for $q_0=0$, 
\begin{equation}
J_0^* = J_0(q,\mu)|_{q_0=0} \equiv J_0^c
\end{equation}
 which corresponds to the case of  zero external field in the associated model and hence $m=0$, 
then 
\begin{equation}
  m \to 0\ \mbox{as}\ J_0 \to {J_0^c}^+
\end{equation}
and the PM/FM transition is {\sl continuous}.

If,  on the contrary, the minimum occurs at a finite value of $q_0$, then $m\not=0$ at  $J_0^*$ 
and the FM phase appears {\sl discontinuously}. In this case the critical point 
$J_0^*$ in general  corresponds to a {\sl spinodal} point, where the solution first appears. 
The true thermodynamic (discontinuous) transition occurs at $J_0^d \geq J_0*$, where the
free energy $\Phi$ of the PM and FM solutions become equal.

\subsection{The $2+p$ Model}
From the form of the unfolding equation (\ref{eq:unfsp}) we see that a continuous transition,
i.e. a finite $J_0^c$, is possible only if $s=2$,  when the first $m$ disappears from the denominator 
so that $b_p$ is finite for $m=0$:
\begin{equation}
\label{eq:unf2p}
   b_p\Bigr|_{m=0} =\frac{1}{2\gamma \chi(0)} 
\end{equation}
leading to
\begin{equation}
  J_0^c = \frac{2\gamma  + p!}{2\gamma}\,\frac{T}{\chi(0)}
\end{equation}
The explicit form of $J_c$ depends on the structure of the solution of the associated problem with 
zero external field $b$.
If for the given temperature $T$ (or $\mu_p$) the $b=0$ phase is Replica Symmetric (RS) then 
$\chi(q_0) = 1 - q_0$, and 
\begin{equation}
  J_0^c = \frac{2\gamma  + p!}{2\gamma}\, T, \qquad \mbox{(RS)}
\end{equation}
In the case of a one-step replica symmetry (1RSB) phase $\chi(q_0) = 1 - q_1 + x(q_1 - q_0)$ 
and
\begin{equation}
 J_0^c = \frac{2\gamma  + p!}{2\gamma}\,\frac{T}{1 - q_1 + x q_1}, \qquad \mbox{(1RSB)}
\end{equation}
Finally if the phase is FRSB or 1FRSB, then $\chi(q_0) = 1/\sqrt{\Lambda'(q_0)}$
and $\chi(0) = 1/\sqrt{\mu_2}$, leading to
\begin{equation}
  J_0^c = \frac{2\gamma  + p!}{2\gamma}\,\sqrt{\frac{pr}{2+pr}}, \qquad \mbox{(FRSB/1FRSB)}
\end{equation}
If for a given $T$, or $\mu_p$,  the minimum of $J_0$ occurs at a physically acceptable finite 
$q_0$ with $J_0^* < J_0^c$, then there is a spinodal point and the transition turns discontinuous.
Besides the boundary values of $q_0$, the minimum of $J_0^*$ may occur at the the stationary 
point of $b_p$: $d b_p/dq_0|_{\mu} = 0$.
A straightforward calculation for the $2+p$ case yields
\begin{eqnarray}
\left.\frac{d b_p}{d q_0}\right|_{\mu} &= & - \frac{1}{(2\gamma + p m^{p-2})^2\chi(q_0)^2}
\\
&&\hspace*{-1cm}
\times
  \biggl\{ 2\gamma \chi'(q_0) 
                + p \bigl[q_0 - (p-1)\chi^2\Lambda(q_0)\bigr]\chi'(q_0) m^{p-4} 
                \nonumber 
                \\
                &&\qquad \qquad\qquad 
                +\frac{p(p-2)}{2}\bigl[1 - \chi^2(q_0) \Lambda'(q_0) \bigr] \chi(q_0) m^{p-4}
  \biggr\}
  =0
  \nonumber
\end{eqnarray}

\noindent where
\begin{equation}
 m^2 = q_0 - \chi(q_0)^2 \Lambda(q_0)
\end{equation}
\begin{figure}[t!] 
   \centering
  \includegraphics[width=.89\textwidth, angle=0]{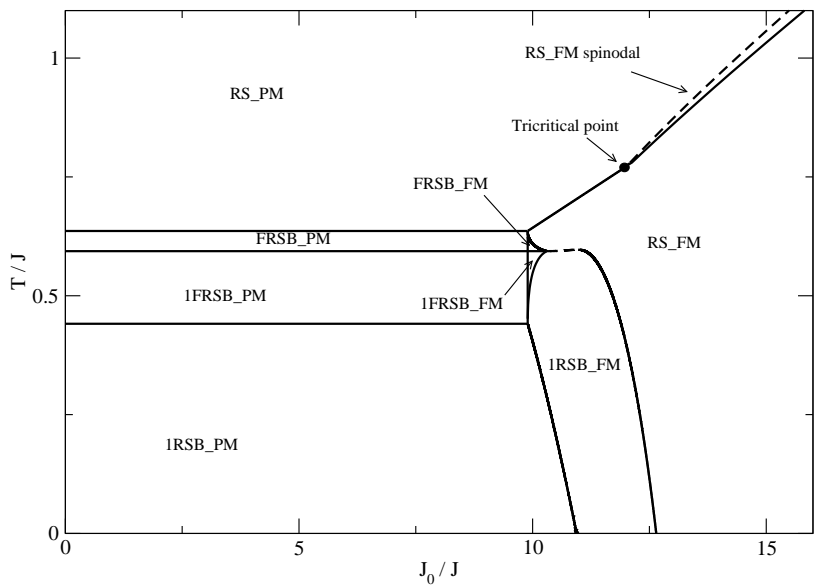} 
   \caption{
   	Phase diagram of the $2+4$ model in the $T$, $J_0$ plane for $r=0.340$ and 
    	$\gamma= 0.8246$. The transition between the PM phases ($m=0$) and the FM phases
	($m\not=0$) is continuous in $m$ up to the tricritical point. Above this point the 
	transition between the paramagnetic phase, denoted RS\_PM, and the ferromagnetic phase, 
	denoted RS\_FM,  is discontinuous with a finite jump of $m$ along the transition line 
	(full line in figure). The transition is accompanied by the appearance of a ferromagnetic 
	spinodal line (dashed line in figure).
           }
   \label{fig:pd24_TJ}
\end{figure}

\begin{figure}[t!] 
   \centering
  \includegraphics[width=.89\textwidth, angle=0]{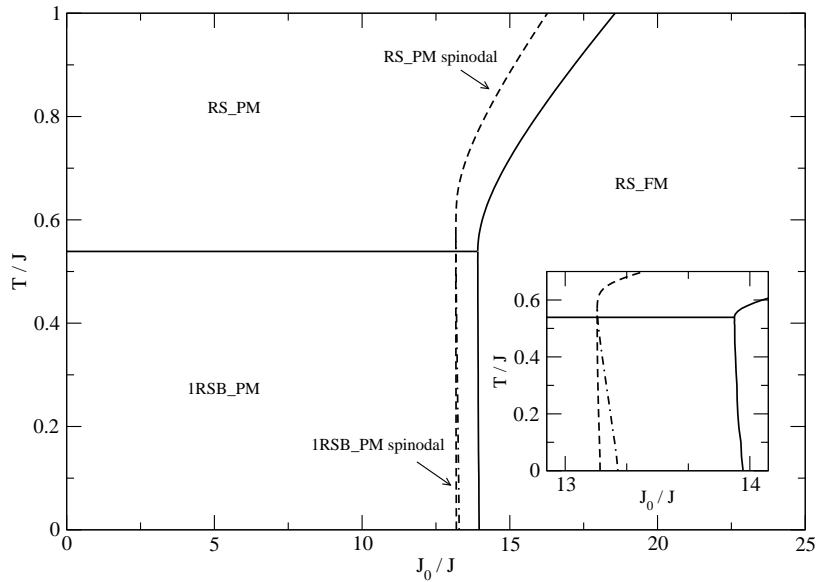} 
   \caption{
   	Phase diagram of the $3+4$ model in the $T$, $J_0$ plane for $r=0.340$ and 
    	$\gamma= 1$. The transition between the PM phases ($m=0$) and the FM phases
	($m\not=0$) is always discontinuous with a finite jump of $m$ along the transition line 
	(full line in figure). The transition is accompanied by the appearance of a ferromagnetic 
	spinodal line (dashed line in figure). The inset shows the transition line 
	(dash-dotted line) between the
	(metastable) ferromagnetic 1RSB phase that appears at the spinodal line (dashed line)
	and the (metastable) ferromagnetic (RS) phase. The latter becomes the thermodynamic
	stable phase at the transition line (full line on the rhs.). 
           }
   \label{fig:pd34_TJ}
\end{figure}
In Fig. \ref{fig:pd24_TJ} we report the phase diagram in the $(T,J_0)$ plane
for $p=4$, $r=0.340$ and $\gamma=0.8246$. Up to some 
temperature dependent threshold value of  the ``ordering parameter" $J_0/J$, 
the phases have $m=0$ (``PM" phases in Figure \ref{fig:pd24_TJ})   and are described by the $b=0$ limit of
the solutions discussed in the previous Sections.  
As $J_0/J$ increases, phases with $m\not=0$ (FM phases in Figure \ref{fig:pd24_TJ}) appear. 
For low enough temperature the transition between the PM phases and the FM phases
occurs continuously with $m$ vanishing at the transition. When the temperature is raised one 
eventually hits a tricritical point where the transition turns discontinuous. Above this temperature 
the transition occurs with a finite jump in $m$ on the transition line (full line in Figure \ref{fig:pd24_TJ}), and is 
accompanied by the presence of a spinodal line where the $m\not=0$ first appears 
(dashed line). 
Interestingly, there exists a range of $J_0$ where, upon cooling, the system goes from a high 
temperature paramagnetic phase to a low temperature 1RSB spin glass phase with $m=0$ 
passing through intermediate FM phases. In this case no continuous transition occurs between the 
PM and the FM phases. This transition is the counterpart of the 1RSB/RS transition that occurs 
in the model with a uniform external field.

When $s>2$ then $b_p\to\infty$ as $m\to0$ and the transition between PM and FM phases 
can  occur only discontinuously with a jump in $m$.\footnote{A similar scenario also 
occurs for $2+p$ models if $\gamma=0$.} As an example in Fig. \ref{fig:pd34_TJ} we show the 
phase diagram of the $3+4$ model for $r=0.75$ and $\gamma=1$. Note the presence of a 
phase transition (dash-dotted line) between the (metastable) 1RSB PM solution, which appears 
at the spinodal line (dashed line),  and the (metastable) RS FM solution (see also inset in figure
\ref{fig:pd34_TJ}).

\section{Conclusions}

In this work we have studied the spherical multi-$p$-spin model with ferromagnetic interactions. 
 We formally add multi $k-$body interaction terms with deterministic interactions next to multi $p$-body 
 terms with quenched disordered couplings of zero average.
  A particular case of this set of interactions is to have
quenched disorder with non-zero average.
After recalling in detail the features of this class of models we have 
shown  that adding purely ferromagnetic terms to
the quenched disordered ones can be simply encoded into
adding an effective field acting on the purely disordered system. 
More specifically, once that the  presence of the ordered part of the
Hamiltonian is encoded into a field, it does not play any role anymore and
one can study the systems properties decoupling the analysis    
of  the behavior of a model in a  field from the computation
of the relationship between the effective field and the real magnetizations
 brought about by the ferromagnetic couplings.
In the replica symmetry breaking parameters, 
 the value of the field only
enters in setting the value of the lowest value $q_0$ of the generic overlap function
$q(x)$.  This is
an increasing function of the field and, hence, the presence
of the ferromagnetic contribution can only reduce the ``complexity"
of the solution found in absence of it. In particular, by increasing
the value of the effective field we can eventually force the {\em frozen} solution to be  a
RS solution. 

We have  detailed the analysis of some specific examples whose properties are, though, general.
The simplest one is the $3+4$ model, whose phase diagram is akin to a single $p$-spin model.
We then show the behavior of the $2+3$ model, still displaying only RS and 1RSB phases (both 
with and without ferromagnetic ordering) but whose transitions can be both discontinuous and 
continuous. On the warm side of  the dynamic transition line this is a realization of the mode 
coupling $F_{12}$ schematic theory \cite{Goetze89b, Goetze89c}.
Eventually we exhaustively describe the behavior of  the $2+p$ model (with $p\geq 4$) where 
many phases of different 
complexity level arise: RS, 1RSB, FRSB and 1FRSB both with and without ferromagnetic ordering. 
In particular,
 a continuous breaking of the replica symmetry is realized at low temperature and field in a given 
 region of the phase space,
 cf. Figs. \ref{fig:pd23_mub}, \ref{fig:pd24_TJ} for the explicit case $p=4$.
 From a dynamic perspective the model is equivalent to a mode coupling $F_{13}$ schematic 
 theory \cite{Goetze89b, Goetze89c}.
We note that the analysis has been performed using the static approach. 
When the phase is described by a step-like order parameter function $q(x)$, such as in the 
1RSB phase, or it possesses a step-like part,  as in the 1FRSB phase, the location of the transition 
between different  phases in the phase space can be different, if one considers the dynamic 
properties of the system.
Roughly speaking this is a consequence of a the presence of a macroscopic number of 
metastable states that prevents the dynamics to reach the lowest (stable) state. The dynamical 
transition takes place at the point where the metastable states become dominant, and 
occurs before the static critical point is reached. 
As far as the phase diagram is concerned the differences between statics and dynamics are 
then mainly quantitative, not qualitative. 
In order not to dull reading with too many details we have not explicitly considered this difference in the main text.
Nevertheless for completeness a brief technical discussion about dynamic equations and 
transition lines has been reported in  \ref{App:B}.

\section*{Acknowledgments}
The authors acknowledge F. Krzakala, Y.F. Sun, and L. Zdeborova for stimulating interaction.
The research leading to these results has received funding
from  the People Programme (Marie Curie Actions) of the European Union's Seventh Framework Programme FP7/2007-2013/ under REA grant agreement n¡ 290038, NETADIS project and 
from the Italian MIUR under the Basic
Research Investigation Fund FIRB2008 program, grant
No. RBFR08M3P4, and under the PRIN2010 program, grant code 2010HXAW77-008.

\appendix
\section{1RSB-1FRSB critical surface for the  $2+p$ model}
\label{App:A}
The 1RSB equations for the $2+p$ model are
\begin{equation}
\label{eq:2p1rsbq0}
  \mu_2 q_0 + \mu_p q_0^{p-1} = \frac{q_0}{\chi(q_0)^2} - b^2
\end{equation}
\begin{equation}
  \mu_2 (q_1 - q_0) + \mu_p (q_1^{p-1} - q_0^{p-1}) = \frac{q_1 - q_0}{\chi(q_1) \chi(q_0)} 
\end{equation}
where
\begin{equation}
  \chi(q_1) = 1 - q_1, \qquad \chi(q_0) = 1 - q_1 + x (q_1 - q_0)
\end{equation}
and the breaking point $x$ is fixed by equation (\ref{eq:1rsbz}).
The 1RSB phase is stable provided the eigenvalue $\Lambda_0^{(3)}$, eq. (\ref{eq:eig03}),
is positive. This leads to the critical condition
\begin{equation}
\label{eq:Acrit}
 \Lambda'(q_0) = \frac{1}{\chi(q_0)^2} \quad\Rightarrow\quad
 \mu_2 + \mu_p(p - 1) q_0^{p-2} = \frac{1}{\chi(q_0)^2}.
\end{equation}
By introducing the ratio $t = q_0 / q_1 \in [0,1]$, and solving the equations for $\mu_2$ and 
$\mu_p$, we obtain the equation of the 1RSB critical line
\begin{eqnarray}
\label{eq:mup1rsb2p}
  \mu_p &=& \frac{1}{x^2 y (1-y)^{p-3} (1-t)}\,
                 \frac{[1 - y + xy (1-t)]^p}{ [1 - (p-1) t^{p-2} + (p-2) t^{p-1}] }
\\
\label{eq:mu21rsb2p}
  \mu_2 &=& \frac{ [ y (1 - t^{p-1}) - (p-1)(1-t)t^{p-2} ]}
                          {x^2 y (1-t)^2}\,
                          \\
                          \nonumber
                          && \hspace*{4 cm}\times
                 \frac{[1 - y + xy (1-t)]^2}
                          { [1 - (p-1) t^{p-2} + (p-2) t^{p-1}] }
\\
\label{eq:b1rsb2p}
   b^2 &=& (p-2) 
                \frac{ t^{p-1} (1-y)^2 }
                        { x^2 y (1-t) }\,
                \frac{ [1 - y + x y (1 - t)] }
                        { [1 - (p-1) t^{p-2} + (p-1) t^{p-1}] }
\end{eqnarray}
where $y$ and $t$ are related by the equation  
\begin{eqnarray}
\label{eq:2pzy}
 z(y) &=& \biggl\{
   y [ p-2 - p t + p t^{p-1} - (p-2) t^p] 
   \\
   \nonumber
   &&\quad+ 2 - p(p-1)t^{p-2} + 2 p(p-2)t^{p-1} - (p-1)(p-2) t^p
   \biggr\}
   \\
   \nonumber
&&   \qquad \times\frac{1} {p (1-t) [ 1 - (p-1) t^{p-2} + (p-2) t^{p-1}]}.
\end{eqnarray}
and 
\begin{equation}
\label{eq:r1rsb2p}
   r =  \frac{\mu_2}{\mu_p}= \frac{ (1-y)^{p-3} }
                    {(1-t)}\,
        \frac{ [ y (1 - t^{p-1}) - (p-1) t^{p-2}] }
                    {[1 - y + x y (1 - t)]^{p-2}  }.
\end{equation}
For the $2+4$ spherical model the equations can be simplified
as follows 
\begin{eqnarray}
  \mu_4 &=& \frac{[1 - y + x y (1 - t)]^4}{(1-t)^3 (1+2t) x^2 y (1 - y)}
\\
  \mu_2 &=& \frac{[y (1 + t + t^2) - 3t^2] [1 - y + x y (1 - t)]^2}{(1-t)^3 (1+2t) x^2 y}
\\
  b^2 &=& 2\, \frac{t^3 (1-y)^2 [1 - y + x y (1-t)]}{(1-t)^3 (1+2 t) x^2 y},
\end{eqnarray}
where $y$ is solution of
\begin{equation}
\label{eq:24zy}
  z(y)  = \frac{1 + 3t + y (1+t)}{2 ( 1 + 2t)},
\end{equation}
and 
\begin{equation}
  r = \frac{\mu_2 }{ \mu_4} = (1-y)\, \frac{y (1 + t + t^2) - 3 t^2}{[1 - y + x y (1 -t)]^2}.
\end{equation}

Equation (\ref{eq:2pzy}), or (\ref{eq:24zy}) for the $2+4$ case, 
can be solved by fixing the value of either $y$ or $t$ 
in the range $[0,1]$, and solving for the other.
Once $x$, $y$ and $t$ are known, $\mu_p$, $\mu_2$ and $b$ are obtained from eqs. (\ref{eq:mup1rsb2p}), (\ref{eq:mu21rsb2p}) and (\ref{eq:b1rsb2p}). The other quantities are
given by
\begin{equation}
 q_1 = \frac{1-y}{1 - y + x y (1 - t)}, \quad
 \chi(q_0) = \frac{x (1-t)}{1 - y + x y (1-t)}
\end{equation}
with $q_0 = t q_1$, $\chi(q_1) = y \chi(q_0)$.

By fixing $x$ and solving for $y$ as function of $t$ one finds the $x$-lines in the $(\mu_p, \mu_2, b)$
space along which the 1RSB phase becomes unstable. These are obtained by setting $t=0$  ($q_0 = 0$) into Eqs.
(\ref{eq:mup1rsb2p}), (\ref{eq:mu21rsb2p}) and (\ref{eq:2pzy}), and are reported in Sec. \ref{sec:2+p},
cf. Eqs. (\ref{eq:mup1rsb2pt0})-(\ref{eq:2pzyt0}).

In Sec. \ref{sec:2+p} we also report the boundary values $r^{(t)}_x$ for which the constant $r$ $(\mu_4,b)$ projections
start displaying different phases, cf. Eqs. (\ref{eq:r_1_1}), (\ref{eq:r_1_0}), (\ref{eq:r_0_0}), (\ref{eq:r_0_1}).

The 1RSB-1FRSB critical surface ends when $t=1$ 
where the discontinuity associated with the 1RSB (and 1FRSB) solution disappears.
Solving equation (\ref{eq:2pzy}) in the limit $t \to 1^{-}$ we obtain
\begin{equation}
 y = 1 - \frac{p-3}{3} (1 - t) + O\left((1-t)^2\right), \qquad t\to 1^-
\end{equation}
and
\begin{equation}
  q_1 = \frac{p - 3}{p - 3(1-x)} + O(1-t), 
  \qquad q_1 - q_0 = O(1 -t),
  \qquad t \to 1^-.
\end{equation}
The equation of the critical end line of the 1RSB-1FRSB critical surface are reported in 
Sec. \ref{sec:2+p}, 
cf. Eqs. (\ref{eq:mup_1rsb_rs})-(\ref{eq:b2_1rsb_rs}).

\section{Dynamic transition}
\label{App:B}
When the solution is described by a step-like order parameter function $q(x)$, such as in the 
1RSB phase, or it possesses a step-like part, 
as in the 1FRSB phase, the location of the transition between different 
phases in the phase space 
can be different, depending one considers the static or dynamic properties of the system.
This is a consequence of  the appearance of a macroscopic number of 
metastable states that prevents the dynamics to reach the lowest (stable) state \cite{Crisanti95,Cavagna98, Cavagna98b,Crisanti03b}. In this case
 the dynamical evolution of the system is dominated by these metastable states 
and the system fails to reach the static critical point. The dynamical transition is associated with
the point where the effect of the metastable states becomes dominant, and occurs before
the static critical point is reached. 

We do not go into the details of a dynamical study of the model,  but rather use a shortcut that 
allows us to recover the dynamical properties  from the replica calculation described
in the main text. This accounts for replacing the stationary condition of the free energy functional 
$G[q,m]$ with respect to variations of the breaking point $x$ (or $x_1$ for the 1FRSB solution) 
by the (simpler) ``marginal condition"
 \begin{equation}
  \label{eq:Bmarg}
   \Lambda_1^{(1)} = -\Lambda'(q_1) + \frac{1}{\chi(q_1)^2} = 0
 \end{equation}
which describes the critical slowing down of the dynamics at the dynamic transition point.
The interested reader can find more details on this in, e.g., Ref. \cite{Crisanti07a}.
Once this replacement has been done, the study of the phase diagram just follows the same 
mainlines  of described in the main text for the static solution.

For example by solving eqs. (\ref{eq:1rsbq1}) and (\ref{eq:Bmarg}) we have for the 
1RSB phase of the $s+p$ model the parametric equations:
\begin{equation}
 \mu_s = \frac{1}{\chi_1^2\chi_0} 
                \frac{ (q_1^{p-1} - q_0^{p-1}) \chi_0 - (p-1)q_1^{p-2}(q_1-q_0)\chi_1}
                        { (s-1)q_1^{s-2}(q_1^{p-1}-q_0^{p-1}) - (p-1) q_1^{p-2}(q_1^{s-1}- q_0^{s-1})}
\end{equation}
\begin{equation}
 \mu_p = \frac{1}{\chi_1^2\chi_0} 
                \frac{ (q_1^{s-1} - q_0^{s-1}) \chi_0 - (s-1)q_1^{s-2}(q_1-q_0)\chi_1}
                        { (p-1)q_1^{p-2}(q_1^{s-1}-q_0^{s-1}) - (s-1) q_1^{s-2}(q_1^{p-1}- q_0^{p-1})}
\end{equation}
where $\chi_{0,1} = \chi(q_{0,1})$, which give $(\mu_s,\mu_p)$ as function of $(q_0,q_1,x)$.
These equations, with $b$ obtained from eq. (\ref{eq:1rsbq0}), give the complete description
of the 1RSB solution in the dynamic approach. 

In the limit $q_1-q_0\to 0$ the stationary condition used in statics and the marginal condition used
in dynamics coincide, so that all ``continuous transition" are unchanged between static and 
dynamics. This is not true for the discontinuous transition where $q_1-q_0$ remains finite.
To obtain the discontinuous transition surface between the RS and 1RSB phase, we take $x=1$ 
in the above equations and  vary $q_1$ and $q_0$. This surface is qualitatively similar to the 
analogous surface discussed for the static, and, indeed, it  joins the continuous transition
surface along the same line discussed in the main text, but it is everywhere else distinct from that.
It crosses the $b=0$ plane along the line
\begin{equation}
 \mu_s = \frac{p-2 - (p-1) q_1}{(p-s) q_1^{s-2}(1-q_1)^2}
\end{equation}
\begin{equation}
 \mu_p = \frac{(s-1) q_1 - (s-2)}{(p-s) q_1^{p-2}(1-q_1)^2}
\end{equation}
which lies on the left hand side of the corresponding static line in the $(\mu_p,\mu_s)$ plane.

The critical surface between the 1RSB and the 1FRSB phases of the $2+p$ model is obtained 
from the above equations by imposing the additional critical condition (\ref{eq:Acrit}), which 
reduces the number of free parameters from $3$ to $2$. A straightforward calculations leads 
again to eqs. (\ref{eq:mup1rsb2p})-(\ref{eq:b1rsb2p}) with 
\begin{equation}
 y = \frac{ 1 - (p-1) t^{p-2} + (p-2) t^{p-1}} {(p-1)(1-t) - (1 - t^{p-1})}
\end{equation}
replacing  eq. (\ref{eq:2pzy}). This surface intersects the continuous 1RSB-RS transition surface
along the same line discussed in the main text for the static,  
eqs. (\ref{eq:mup_1rsb_rs})-(\ref{eq:b2_1rsb_rs}), and the $b=0$ plane on the line of equation
(\ref{eq:mup1rsb2pt0})-(\ref{eq:mu21rsb2pt0}) with $y_0 = 1/(p-2)$.
 The analysis for other cases is straightforward.



\section*{References}
 \begin{thebibliography}{10}
\expandafter\ifx\csname url\endcsname\relax
  \def\url#1{\texttt{#1}}\fi
\expandafter\ifx\csname urlprefix\endcsname\relax\def\urlprefix{URL }\fi
\expandafter\ifx\csname href\endcsname\relax
  \def\href#1#2{#2} \def\path#1{#1}\fi

\bibitem{Berlin52}
T.~H. Berlin, M.~Kac, {\em{The spherical model of a ferromagnet}}, Phys. Rev.
  {\bf 86} (1952) 821Ð835.

\bibitem{Kirkpatrick87b}
T.~Kirkpatrick, D.~Thirumalai, \em{$p$-spin-interaction spin-glass models:
  Connections with the structural glass problem}\em, Phys. Rev. B {\bf 36}
  (1987) 5388.

\bibitem{Kirkpatrick87c}
T.~Kirkpatrick, D.~Thirumalai, \em{Dynamics of the structural glass transition
  and the p-spin interaction spin-glass model}\em, Phys. Rev. Lett. {\bf 58}
  (1987) 2091.

\bibitem{Thirumalai88}
D.~Thirumalai, T.~Kirkpatrick, \em{Mean-field Potts glass model:
  Initial-condition effects on dynamics and properties of metastable
  states}\em, Phys. Rev. B {\bf 38} (1988) 4881.

\bibitem{Crisanti92}
A.~Crisanti, H.~Sommers, \em{The spherical $p$-spin interaction spin-glass
  model - the statics}\em, Z. Phys. B {\bf 87} (1992) 341.

\bibitem{Crisanti07b}
A.~Crisanti, L.~Leuzzi, \em{Amorphous-amorphous transition and the two-step
  replica symmetry breaking phase}\em, Phys. Rev. B {\bf 76} (2007) 184417.

\bibitem{Crisanti11}
A.~Crisanti, L.~Leuzzi, M.~Paoluzzi, \em{Statistical mechanical approach to
  secondary processes and structural relaxation in glasses and glass
  formers}\em, Eur. Phys. J. E {\bf 34} (2011) 98.

\bibitem{Romanini12}
M.~Romanini, P.~Negrier, J.~L. Tamarit, S.~Capaccioli, M.~Barrio, L.~C. Pardo,
  D.~Mondieig, {\em Emergence of glassy-like dynamics in an orientationally
  ordered phase}, Phys. Rev. B 85 (2012) 134201.

\bibitem{Ngai11}
K.~L. Ngai, \em Relaxation and Diffusion in Complex Systems \em, Springer
  Verlag (New York), 2011.

\bibitem{Goetze89b}
W.~G{\"o}tze, L.~Sj{\"o}gren, {\em Beta relaxation near glass transition
  singularities }, J. Phys.: Cond. Matt. 1 (1989) 4183.

\bibitem{Goetze89c}
W.~G{\"o}tze, L.~Sj{\"o}gren, {\em Logarithmic decay laws in glassy systems },
  J. Phys.: Cond. Matt. 1 (1989) 4203.

\bibitem{Goetze09}
W.~{G\"otze}, \em{Complex Dynamics of Glass-Forming Liquids: A Mode-Coupling
  Theory}\em, OUP (Oxford, UK), 2009.

\bibitem{Mezard91}
M.~M{\'e}zard, G.~Parisi, {\em Replica field theory for random manifolds}, J.
  Phys. I {\bf 1} (1991) 809.

\bibitem{Giamarchi94}
T.~Giamarchi, P.~Le~Doussal, {\em Elastic theory of pinned flux lattices},
  Phys. Rev. Lett. {\bf 72} (1994) 1530.

\bibitem{Giamarchi95}
T.~Giamarchi, P.~Le~Doussal, {\em Elastic theory of flux lattices in the
  presence of weak disorder}, Phys. Rev. B {\bf 52} (1995) 1242.

\bibitem{Cugliandolo96}
L.~F. Cugliandolo, J.~Kurchan, P.~Le~Doussal, {\em Large Time
  Out-of-Equilibrium Dynamics of a Manifold in a Random Potential}, Phys. Rev.
  Lett. {\bf 76} (1996) 2390.

\bibitem{LeDoussal98}
P.~Le~Doussal, K.~J. Wiese, {\em Glassy Trapping of Manifolds in Nonpotential
  Random Flows}, Phys. Rev. Lett. {\bf 80} (1998) 2362.

\bibitem{Nieuwenhuizen95}
T.~Nieuwenhuizen, \em{To maximize or not to maximize the free energy of glassy
  systems}\em, Phys. Rev. Lett. {\bf 74} (1995) 3463.

\bibitem{Crisanti04b}
A.~Crisanti, L.~Leuzzi, \em{Spherical $2+p$ spin-glass model: An exactly
  solvable model for glass to spin-glass transition}\em, Phys. Rev. Lett. {\bf
  93} (2004) 217203.

\bibitem{Crisanti06}
A.~Crisanti, L.~Leuzzi, \em{Spherical $2+p$ spin-glass model: An analytically
  solvable model with a glass-to-glass transition}\em, Phys. Rev. B {\bf 73}
  (2006) 014412.

\bibitem{Sherrington75}
D.~Sherrington, S.~Kirkpatrick, {\em Solvable Model of a Spin-Glass}, Phys.
  Rev. Lett. {\bf 35} (1975) 1792.

\bibitem{Nishimori11}
H.~Nishimori, \em Statistical Physics of Spin Glasses and Information
  Processing: An Introduction \em, Oxford University Press (Oxford), 2001.

\bibitem{Krzakala11}
F.~Krzakala, L.~Zdeborov\'a, {\em{On melting dynamics and the glass transition.
  II. Glassy dynamics as a melting process}\em}, J. Chem. Phys. {\bf 134}
  (2011) 034513.

\bibitem{Barrat97}
A.~Barrat, S.~Franz, G.~Parisi, {\em Temperature evolution and bifurcations of
  metastable states in mean-field spin glasses, with connections with
  structural glasses}, J. Phys. A: Math. Gen. {\bf 30} (1997) 5593.

\bibitem{Capone06}
B.~Capone, T.~Castellani, I.~Giardina, F.~Ricci-Tersenghi, {\em Off-equilibrium
  confined dynamics in a glassy system with level-crossing states}, Phys. Rev.
  B {\bf 74} (2006) 144301.

\bibitem{Sun12}
Y.~F. Sun, A.~Crisanti, F.~Krzakala, L.~Leuzzi, L.~Zdeborov‡, {\em Following
  states in temperature in the spherical s + p-spin glass model}, J. Stat.
  Mech. (2012) P07002.

\bibitem{Thalmann00}
F.~Thalmann, C.~Dasgupta, D.~Feinberg, {\em Phase diagram of a classical fluid
  in a quenched random potential}, Europhys. Lett. {\bf 50} (2000) 54Ð60.

\bibitem{Krakoviack07}
V.~Krakoviack, {\em Mode-coupling theory for the slow collective dynamics of
  fluids adsorbed in disordered porous media}, Phys Rev E {\bf 75} (2007)
  031503.

\bibitem{Krakoviack10}
V.~Krakoviack, {\em Statistical mechanics of homogeneous partly pinned fluid
  systems}, Phys Rev E {\bf 82} (2010) 061501.

\bibitem{Gordon02}
A.~Gordon, B.~Fischer, {\em Phase Transition Theory of Many-Mode Ordering and
  Pulse Formation in Lasers}, Phys. Rev. Lett. 89 (2002) 103901.

\bibitem{Gordon03}
A.~Gordon, B.~Fischer, {\em Phase transistion theory of pulse formation in
  passively mode-locked lasers with dispersion and Kerr nonlinearity}, Opt.
  Comm. 223 (2003) 151--156.

\bibitem{Weill05}
R.~Weill, A.~Rosen, A.~Gordon, O.~Gat, B.~Fischer, {\em Critical Behavior of
  Light in Mode-Locked Lasers}, Phys. Rev. Lett. 95~(1) (2005) 013903.

\bibitem{Cao99}
H.~Cao, Y.~G. Zhao, S.~T. Ho, E.~W. Seelig, Q.~H. Wang, R.~P.~H. Chang, {\em
  Random Laser Action in Semiconductor Powder}, Phys. Rev. Lett. 82 (1999)
  2278.

\bibitem{Wiersma08}
D.~S. Wiersma, {\em The physics and applications of random lasers}, Nature
  Physics 4 (2008) 359.

\bibitem{Leuzzi09b}
L.~Leuzzi, C.~Conti, V.~Folli, L.~Angelani, G.~Ruocco, {\em Phase Diagram and
  Complexity of Mode-Locked Lasers: From Order to Disorder}, {Phys. Rev. Lett.}
  {{\bf 102}} ({2009}) 083901.

\bibitem{Conti11}
C.~Conti, L.~Leuzzi, {\em Complexity of waves in nonlinear disordered media },
  Phys. Rev. B {\bf 83} (2011) 134204.

\bibitem{Crisanti93}
A.~Crisanti, H.~Horner, H.~Sommers, \em{The spherical $p$-spin interaction
  spin-glass model - the dynamics}\em, Z. Phys. B {\bf 92} (1993) 257.

\bibitem{Crisanti07a}
A.~Crisanti, L.~Leuzzi, \em{Equilibrium Dynamics of Spin-Glass Systems}\em,
  Phys. Rev. B {\bf 75} (2007) 144301.

\bibitem{Hertz99}
J.~Hertz, D.~Sherrington, T.~Nieuwenhuizen, {\em Competition between glassiness
  and order in a multispin glass}, Phys. Rev. E {\bf 60} (1999) R2460--3.

\bibitem{Krakoviack07b}
V.~Krakoviack, Comment on ``spherical $2+p$ spin-glass model: An analytically
  solvable model with a glass-to-glass transition'', Phys. Rev. B 76 (2007)
  136401.

\bibitem{Crisanti07c}
A.~Crisanti, L.~Leuzzi, Reply to ``comment on `spherical $2+p$ spin-glass
  model: An analytically solvable model with a glass-to-glass transition' '',
  Phys. Rev. B 76 (2007) 136402.

\bibitem{Mezard87}
M.~M\'ezard, G.~Parisi, M.~Virasoro, Spin Glass Theory and Beyond, World
  Scientific (Singapore), 1987.

\bibitem{Parisi79a}
G.~Parisi, {\em Infinite Number of Order Parameters for Spin-Glasses}, Phys.
  Rev. Lett. {\bf 43} (1979) 1754Ð1756.

\bibitem{Parisi79b}
G.~Parisi, {\em Toward a mean field theory for spin glasses}, Phys. Lett. A
  {\bf 73} (1979) 203--205.

\bibitem{Parisi80}
G.~Parisi, \em{A sequence of approximated solutiona to the S-K model for spin
  glasses}\em, \em J. Phys. A: Math. Gen.\em {\bf 13} (1980) L115.

\bibitem{Carlucci96}
D.~M. Carlucci, C.~De~Dominicis, T.~Temesvari, {\em Stability of the
  MŽzard-Parisi Solution for Random Manifolds}, J. Phys. I (France) {\bf 6}
  (1996) 1031.

\bibitem{DeDominicis97}
C.~De~Dominicis, D.~M. Carlucci, T.~Temesvari, {\em Replica Fourier Tansforms
  on Ultrametric Trees, and Block-Diagonalizing Multi-Replica Matrices}, J.
  Phys. I (France) {\bf 7} (1997) 105--115.

\bibitem{Crisanti04}
A.~Crisanti, F.~Ritort, {\em{Intermittency of glassy relaxation and the
  emergence of a non-equilibirum spontaneous measure in the aging regime}\em},
  {Europhys. Lett.} {\bf 66} (2004) 253.

\bibitem{Crisanti95}
A.~Crisanti, H.~Sommers, {\em{Thouless-Anderson-Palmer approach to the
  spherical p-spin spin glass model}\em}, J. Phys. I (France) {\bf 5} (1995)
  805--813.

\bibitem{Cavagna98}
A.~Cavagna, I.~Giardina, G.~Parisi, \em{Stationary points of the
  Thouless-Anderson-Palmer free energy}\em, Phys. Rev. B {\bf 57} (1998) 11251.

\bibitem{Cavagna98b}
A.~Cavagna, J.~P. Garrahan, I.~Giardina, {\em Quenched complexity of the
  mean-field p-spin spherical model with external magnetic field}, J. Phys. A
  {\bf 32} (1998) 711.

\bibitem{Crisanti03b}
A.~Crisanti, L.~Leuzzi, T.~Rizzo, {\em{The Complexity of the Spherical $p$-spin
  spin glass model, revisited}}, Eur. Phys. J. B {\bf 36} (2003) 129--136.

\end{thebibliography}

\end{document}